\documentclass[12pt]{article}
\usepackage[top=1in, bottom=0.75in, left=0.75in, right=0.75in]{geometry}
\usepackage{graphicx}
\usepackage{amssymb}
\usepackage{algorithm}
\usepackage{algorithmic}
\usepackage{amsmath}
\usepackage{amsfonts}
\usepackage{dsfont}
\usepackage{url}
\usepackage{chemarr}
\usepackage{chemarrow}
\usepackage{bbold}
\usepackage[version=3]{mhchem}
\usepackage{mathabx}

\newtheorem{thm}{Theorem}
\newtheorem{example}{Example}

\begin{document} 

\title{Impact of receiver reaction mechanisms on the performance of molecular communication networks} 
\author{Chun Tung Chou \\
School of Computer Science and Engineering \\ University of New South Wales \\ Sydney, NSW 2052, Australia \\
E-mail: ctchou@cse.unsw.edu.au}



\maketitle

\begin{abstract}
In a molecular communication network, transmitters and receivers communicate by using signalling molecules. At the receivers, the signalling molecules react, via a chain of chemical reactions, to produce output molecules. The counts of output molecules over time is considered to be the output signal of the receiver. This output signal is used to detect the presence of signalling molecules at the receiver. The output signal is noisy due to the stochastic nature of diffusion and chemical reactions. The aim of this paper is to characterise the properties of the output signals for two types of receivers, which are based on two different types of reaction mechanisms. We derive analytical expressions for the mean, variance and frequency properties of these two types of receivers. These expressions allow us to study the properties of these two types of receivers. In addition, our model allows us to study the effect of the diffusibility of the receiver membrane on the performance of the receivers. 
\end{abstract} 


\section{Introduction} 
\label{sec:intro} 
A molecular communication network \cite{Akyildiz:2008vt,Hiyama:2010jf,Nakano:2012dv} consists of multiple transmitters and receivers in a fluid medium. The transmitters encode the messages by types, concentration or emission frequency of signalling molecules. The signalling molecules diffuse freely in the medium. When these signalling molecules reach the receiver, they trigger chemical reactions in the receiver to allow their presence to be detected. There are a number of reasons why the study of molecular communication networks, both natural and synthetic, are important. First, molecular communication is a fundamental ingredient of life on earth. Multi-cellular organisms make extensive use of molecular communication to regulate body functions \cite{Alberts}. Second, synthetic molecular communication networks can be used as sensor networks for cancer detection and treatment \cite{Atakan:2012ej}, as well as many other applications \cite{Nakano:2012dv}. 

An important part of the research on molecular communication networks is to understand their performance. Work done in this area includes the modelling of molecular communication networks \cite{Pierobon:2010vg,Chou:rdmex_tnb}, capacity analysis \cite{Atakan:2010bj,Pierobon:2013cl,Srinivas:2012et}, characterising the noise of transmitters, channels and receivers \cite{Pierobon:2011vr,Pierobon:2011ve,Mahfuz:2011te,Miorandi:2011jw,Leeson:2012uy,Chou:rdmex_nc}, and receiver design \cite{Chou:2012ug,ShahMohammadian:2013jm,ShahMohammadian:2012iu,Noel:2013tr}. This paper will focus on the receiver part of the network.

When signalling molecules arrive at the receiver, they react, via one or more chemical reactions, to produce a number of output molecules. The number (or counts) of output molecules in the receiver over time is regarded as the {\sl output signal} of the receiver. A way to detect the presence of signalling molecules at the receiver is to test whether the number of output molecules at the receiver is more than a threshold. However, the output signal is {\sl noisy} due to the stochastic nature of diffusion and chemical reactions. An important part of performance analysis is to characterise the probability distribution, mean and variance of the receiver output signals. The aim of this paper is to characterise two different types of receiver reaction mechanisms, namely {\sl reversible conversion} and {\sl linear catalytic}, and compare their performance. The main contributions of this paper are:
\begin{itemize}
\item We characterise the output signals of receivers that use reversible conversion in terms of probability distribution, mean, variance and frequency response. 
\item We derive the mean, variance and frequency response of the output signal for receivers that use the linear catalytic reaction mechanism. 
\item We investigate the properties of the output signals of these receivers. For reversible conversion, we find that the variance of the output signal is always smaller than the mean output signal. However, the linear catalytic reaction magnifies the noise to create an output signal whose variance is greater than the mean. We also find that, for both reaction mechanisms, an increase in mean output also leads to an increase in variance. 
\item We show that the diffusibility of the receiver membrane (a parameter which does not appear to have been studied in the literature) can be used to affect the properties of the receiver output signal. 
\end{itemize} 

The rest of the paper is organised as follows. We present our modelling framework in Section \ref{sec:model}. This is followed by performance analysis of the two different types of receiver reaction mechanisms in Sections \ref{sec:sol} and \ref{sec:sol2}. Section \ref{sec:num} presents numerical results. Related work is discussed in section \ref{sec:related}. Finally, Section \ref{sec:con} concludes the paper. 


\section{Modelling molecular communication networks} 
\label{sec:model} 
This section presents a model for molecular communication networks. The model assumes the medium (or space) is discretised into voxels while time is continuous. Features included in the model include geometric constraints, directional emissions of signalling molecules and others. In particular, we will consider two different types of receivers based on different types of chemical reactions. The proposed model will be analysed in Sections \ref{sec:sol} and \ref{sec:sol2}. We will present the {\sl basic} model in Section \ref{sec:model_basic} and various {\sl extensions} will be discussed in Section \ref{sec:model_ext}. 

\subsection{Basic model: assumptions and features}
\label{sec:model_basic} 
A molecular communication network consists of multiple transmitters and receivers. In this paper, we limit ourselves to one transmitter and one receiver. The model and the solution method can be readily generalised to the multiple transmitter and receiver scenario. We assume the transmitter uses one type of signalling molecules $L$. Generalisation to multiple types of non-interacting signalling molecules is straightforward. 

\subsubsection{Transmission medium} 
\begin{figure}
\begin{center}
\includegraphics[width=10cm]{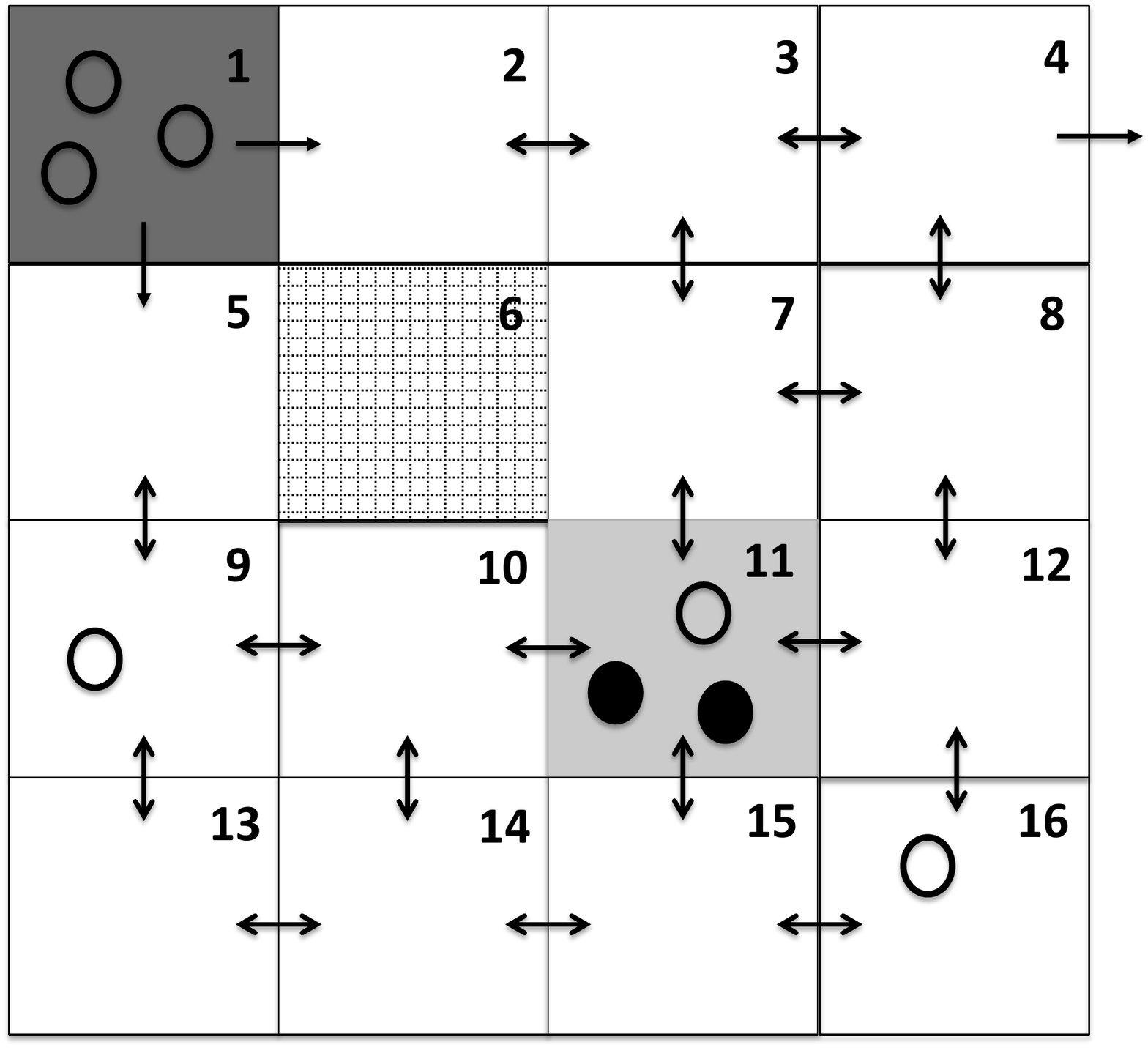}
\caption{A model of molecular communication network. The volume is divided into voxels. The indices of the voxels are given in the top right hand corner. The arrows show when diffusion is allowed. Unfilled circles are signalling molecules. Filled circles are output molecules.}
\label{fig:model}
\end{center}
\end{figure}

We model the transmission medium as a three dimensional (3-D) space with dimensions $X \times Y \times Z$, where $X$, $Y$ and $Z$ are integral multiples of length $\Delta$. That is, there exist positive integers $N_x$, $N_y$ and $N_z$ such that $X = N_x \Delta$ and $Y = N_y \Delta$, $Z = N_z \Delta$. The 3-D volume can be partitioned into $N_x \times N_y \times N_y$ cubic {\sl voxels} of volume $\Delta^3$. Figure \ref{fig:model} shows an arrangement with $N_x = N_y = 4$ and $N_z = 1$.

We refer to a voxel by a triple $(x,y,z)$ where $x$, $y$ and $z$ are integers or by a single index $\xi \in [1,N_x N_y N_z]$ where $\xi(x,y,z) = x + N_x (y-1) + N_x N_y (z-1)$. The indices for the voxels are shown in Figure \ref{fig:model}. 

Diffusion is modelled by molecules moving from one voxel to another. Diffusion from a voxel to a non-neighbouring voxel is always not allowed. The diffusion from a voxel to a neighbouring voxel may or may not be allowed. This can be used to specify different modelling constraints. We use a few examples in Figure \ref{fig:model} to explain this: 
\begin{enumerate}
\item For voxel 11, the diffusion of signalling molecules $L$ is allowed in both directions, i.e. in and out of the voxel. The two-way arrows are used to indicate this. 
\item For voxel 1, signalling molecules can only diffuse out of the voxel but not into it. This is indicated by the one-way arrows. 
\item With the exception of voxel 4, diffusion to the outside of the medium is not allowed. For voxel 4, diffusion to the outside of the medium is allowed for one surface as indicated by the one-way arrow. Our model can be used to capture standard boundary conditions such as reflecting and absorbing boundaries. 
\item No molecules are allowed to flow in and out of voxel 8 (the hatched voxel). This can be used to model shadowing effects or regions that are impermeable to the signalling molecules. 
\end{enumerate}

The {\sl basic} model assumes that the medium is homogeneous with the diffusion coefficient for $L$ in the medium is $D$. Define $d = \frac{D}{\Delta^2}$. If a molecule is allowed to diffuse from a voxel to another, it takes place at a rate of $d$, i.e. within an infinitesimal time $\delta t$, the probability that a molecule diffuses to a neighbouring voxel is $d \delta t$. 

\subsubsection{Transmitters} 
\label{sec:model:transmitters}
We assume the transmitter and the receiver each occupies a distinct voxel. However, it is straightforward to generalise to the case where a transmitter or a receiver occupies multiple voxels. The transmitter (resp. receiver) is assumed to be located at the $T$-th ($R$-th) voxel, where $T$ ($R$) is an integer to indicate the position of the voxel. In Figure \ref{fig:model}, we assume voxel 1 (dark grey) contains the transmitter, and voxel 11 (light grey) contains a receiver. Hence $T = 1$ and $R = 11$ in this example. 

We model a transmitter by a sequence which specifies the number of signalling molecules emitted by the transmitter at a certain time. We assume that, at time $t_i$ (where $i = 1,2,...$), the transmitter emits $k_i$ signalling molecules. This means that: $k_{1}$ signalling molecules are added to the voxel containing the transmitter at time $t_1$, $k_{2}$ signalling molecules are added at time $t_2$ etc. 


Note that in Figure \ref{fig:model}, we only allow signalling molecules to leave the transmitter (voxel 1) but our model can also deal with one-way or two-way flow on any surface of a voxel. 

\subsubsection{Receivers} 
\label{sec:model:rec} 
When a signalling molecule $L$ arrives at a receiver, it may react, via one or more chemical reactions, to produce one or more {\sl output molecules}. We assume that these reactions can only take place within a receiver voxel, and the output molecules do not leave the receiver voxel. In this paper, we consider the following two types of reactions at the receivers.
\begin{enumerate} 
\item The signalling molecules $L$ are converted to complexes $C$, reversibly, at the receiver via the reaction: 
\begin{align}
L \xrightleftharpoons[\text{k}_{-}]{\text{K}_{+}} C 
\label{cr:conversion} 
\end{align}
where $K_+$ and $k_-$ are, respectively, the macroscopic rate constant for the forward and reverse reactions. We will refer to this reaction as {\em reversible conversion} (RC). The output molecule is the complex $C$. 
\item The signalling molecule $L$ acts as a catalyst to produce the output molecule $E$. The molecule $E$ decays at a certain rate. We will refer to this type of receiver as {\sl linear catalytic} or CAT. The chemical formulas for CAT are:
\begin{align}
\cee{
L &  ->[\text{F}_{+}] L + E \label{cr:cat1}  \\
E & ->[\text{f}_{-}] \emptyset 
}
\label{cr:cat2} 
\end{align}
where $F_+$ and $f_-$ are macroscopic rate constants, and $\emptyset$ denotes a chemical species that we are not interested in. 
\end{enumerate} 

For both RC and CAT, the output molecules are formed at a rate of $k_+ = \frac{K_+}{\Delta^3}$ or  $f_+ = \frac{F_+}{\Delta^3}$ times the number of signalling molecules in the voxel. If $k_+ = f_+$ and $k_- = f_-$, then the relation between the mean number of signalling molecules in the voxel and mean number of output molecules is described by exactly the same first order ordinary differential equations for both receiver types. However, as we shall later, the behaviour of the mean and variance of the number of output molecules, for the same transmitter emission pattern, are very different for the two types of receivers. 

Note that both RC and CAT consists of {\sl linear reactions}, which means the reaction rate is a linear function of the concentration of the reactants. Reactions \eqref{cr:conversion} and \eqref{cr:cat2} are examples of {\sl monomolecular} reactions because there is exactly one chemical species on each side of the chemical formula. 

It is appropriate to make a few remarks here. First, RC and CAT can be used to approximate more complex reactions. For example, we show in  \cite{Chou:rdmex_tnb} that RC can approximate more complicated reactions such as Michaelis-Menten. CAT reactions have been studied in \cite{Warren:2006ky}
as a detection mechanism. Second, a receiver may consist of a chain of reactions but we know from data processing inequality \cite{Cover} that the capacity of the network is upper bounded by the output of the first reaction in the chain, therefore we limit our consideration to one reaction.

\subsubsection{Network state and problem statement} 
We define the state of the molecular communication network as the number of signalling molecules in each voxel, the number of complexes in the receiver and the total number of molecules that have left or degraded in the network. Let $n_{L,\xi}(t)$ be the number of signalling molecules in voxel with index $\xi$ at time $t$, $n_{LA}(t)$ be the number of signalling molecules that have either left the medium or degraded by time $t$ and $n_{R}(t)$ be the number of output molecules in the receiver at time $t$. Also let $N_v=N_x N_y N_z$ be the total number of voxels in the medium. The state $N(t) \in {\mathbb Z}^{N_v+2}$ of the network is defined as:
\begin{align}
N(t) = 
\left[ \begin{array}{cccccc}
N_{L,1}(t) & ... & n_{L,N_v}(t)  & n_{LA}(t) & n_{R}(t)  
\end{array}
\right]^T
\label{eqn:state}
\end{align} 
where the superscript $^T$ is used to denote matrix transpose. Note that we also use $T$ and its subscripted form $_T$ to indicate the index of the transmitter voxel. Although the same symbol $T$ is used in different situations, its meaning can easily be deduced from its context. 

We use the network depicted in figure \ref{fig:model} as an example. The unfilled and solid circles represent, respectively, signalling and output molecules. The state vector $N(t)$ has 5 non-zero elements: $n_{L,1}(t) = 3$,  $n_{L,9}(t) = 1$, $n_{L,11}(t) = 1$, $n_{L,16}(t) = 1$ and $n_{R}(t) = 2$. All other elements of $N(t)$ are zero. 

Note that, for $1 \leq i \leq N_v$, the $i$-th element of $N(t)$ is the number of signalling molecules in the $i$-th voxel. A convenient way to represent the emission of molecules by the transmitter is to use the following notation: Assume that at time $t$, $k$ molecules are emitted by the transmitter, which is located at the $T$-th voxel, we can write: $N(t+) = N(t-) + \mathds{1}_{T} k$ where $t-$ and $t+$ are just before and after time $t$, and $\mathds{1}_{T}$ is the $T$-th standard basis vector. For this reason, we also include unreachable voxels, such as voxel 6 in Figure \ref{fig:model}, in $N(t)$ to simplify notation. 

The state $N(t)$ is in fact a vector-valued random process. The state evolves because of diffusion and reactions. We show in \cite{Chou:rdmex_tnb} and \cite{Chou:rdmex_nc} that the probability distribution of $N(t)$ evolves according to a reaction-diffusion master equation with exogenous input (RDMEX), which is a generalisation of the reaction-diffusion master equation (RDME) taking into consideration the emission of signalling molecules by the transmitters. In \cite{Chou:rdmex_tnb}, we show how the mean of $N(t)$ can be calculated for an infinite transmission medium. In \cite{Chou:rdmex_nc}, we show how the covariance of $N(t)$ can be approximately calculated for an infinite medium. We will derive the probability distribution of $N(t)$ for RC receivers in Section \ref{sec:sol}. For CAT receivers, we derive the mean and variance of the number of output molecules in Section \ref{sec:sol2}. 
 
\subsection{Extensions to the basic models} 
\label{sec:model_ext}
\label{sec:ext} 
In this section, we present a number of extensions to the basic model presented earlier. The solution method in Sections \ref{sec:sol} and \ref{sec:sol2} can still be applied to these extensions. 
\begin{enumerate}
\item We assume that each voxel has the shape of a cube. It is possible to use other shapes for the voxel too. However, the rate of diffusion has to be appropriately adjusted according to the geometry, see \cite{Isaacson:2007ui} on how this can be done. 
\item We assume a homogenous medium. An heterogeneous medium with different diffusion constants at different surfaces of the voxels can easily be incorporated. We will consider in later section how heterogeneity can be used to improve the communication performance. 
\item We assume that no flows exist in the medium. Flows can be easy incorporated by adjusting the rate of flow to neighbouring voxels. For example, let us assume there is a flow in the positive $x$ directions of speed $v$, then the rate of flow of molecules from a voxel to its neighbour in the positive $x$ direction should be changed to $\frac{v}{\Delta} + d$. 
\item Degradation of signalling molecules in the voxels can also be handled, in fact in the same way as a molecule leaving the medium. 
\item Multiple transmitters and multiple receivers can be used. 
\item Instead of assuming the number of molecules emitted by the transmitter at a time instance is deterministic, a stochastic emission model can also be handled. 
\end{enumerate}

\section{RC receivers} 
\label{sec:sol}
In this section, we solve for the probability distribution of the state $N(t)$ of a network consisting of one transmitter and one RC receiver. The solution uses the method introduced in \cite{Jahnke:2007th} for solving a system of monomolecular reactions. The reactions \eqref{cr:conversion} in the RC receiver are certainly monomolecular. The diffusion of signalling molecules between the voxels can be viewed as monomolecular reactions if we identify the signalling molecules in each voxel as a distinct chemical species. Therefore, we can view our network as a system of monomolecular reactions. Note that the work in \cite{Jahnke:2007th} does {\sl not} consider transmitters. Our contribution is to extend the method in \cite{Jahnke:2007th} so that it can used to analyse the performance of molecular communication networks. 

An important result for a system of monomolecular reactions is that each molecule evolves {\bf independently} \cite[p.13]{Jahnke:2007th}. This means we can analyse one molecule at a time and ``sum" the results up. We begin with some background. 

\subsection{Background} 
This section is divided into two parts. We first consider Markov chains with one molecule and then we present some background on multinomial distributions. 

\subsubsection{Markov chain model} 
\label{sec:rc:bg:1mc}
Since each molecule in a RC receiver network evolves independently, we present a {\sl continuous-time} Markov chain representation of such networks with only {\sl one} molecule. If there is only one molecule in the network, the molecule can be a signalling molecule in a voxel, in an absorbing state or as a complex in the receiver. We will refer to this Markov chain as the single-molecule Markov chain associated with the network or 1-MC for short.  

\begin{figure}
\begin{center}
\includegraphics[width=10cm]{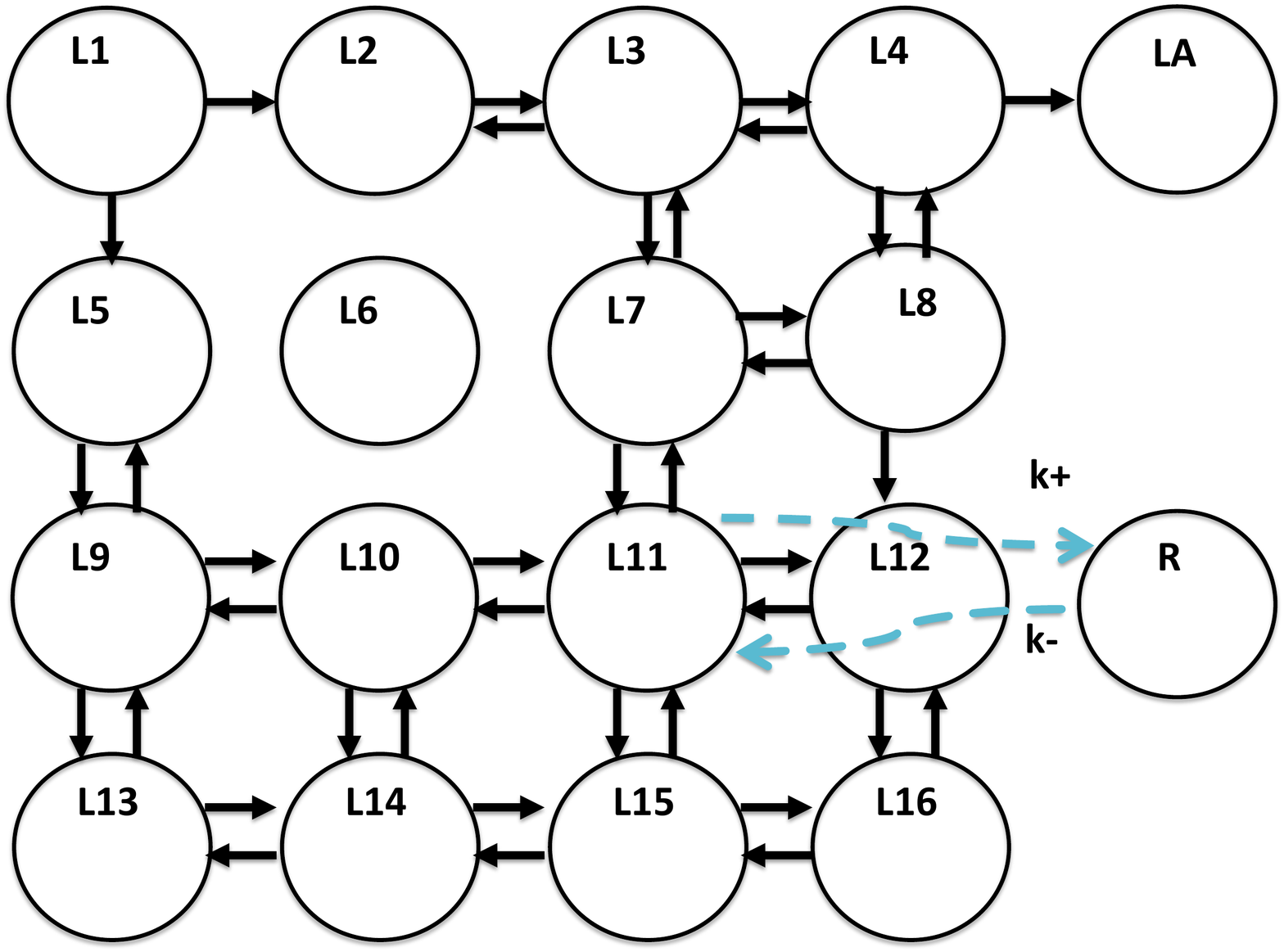}
\caption{The 1-MC for the model shown in Figure \ref{fig:model}, assuming RC receivers.}
\label{fig:mc}
\end{center}
\end{figure}

The easiest way to describe 1-MC is via an example. For the network depicted in Figure \ref{fig:model}, the corresponding 1-MC is depicted in Figure \ref{fig:mc}. In the figure, each circle is a state of the Markov chain. Each state is labelled by $L_i$ ($i = 1,...,16$), $LA$ or $R$. The molecule is in state $L_i$  if it is a signalling molecule in voxel $i$. If the molecule in state $LA$, the molecule has either left the medium or degraded, and is in an absorbing state. If the molecule is in $R$, then it is a complex in the receiver. All the state transitions given in solid lines are due to diffusion and occur at a rate of $d$ for the basic model which assumes homogeneous medium. These rates are not indicated in the figure to avoid cluttering. The conversion rates ($k_+$ and $k_-$) between a signalling molecule and a complex in a receiver are shown in the diagram. 

We will order the states of the 1-MC in the same order as that in the network state vector $N(t)$ in equation \eqref{eqn:state}. Let $q_{L,i}(t)$, $q_{LA}(t)$ and $q_{R}(t)$ be the probability that the molecule is, respectively, a signalling molecule in $i$-th voxel, in absorbing state and in the receiver as a complex. Define the state probability vector $q(t)$ as:
\begin{align}
q(t) = 
\left[ \begin{array}{ccccc}
q_{L,1}(t) & ... & q_{L,N_v}(t)  & q_{LA}(t) & q_{R}(t)   
\end{array}
\right]^T
\label{eqn:qvec}
\end{align} 
The probability vector $q(t)$ of 1-MC evolves according to:
\begin{align}
\dot{q}(t) = H q(t) 
\label{eq:mc1} 
\end{align}
where $H$ is the infinitesimal generator of the 1-MC. The matrix $H$ can be transcribed from the state transition diagram such as Figure \ref{fig:mc}. Note that, when there is only one molecule in the network, there is a one-to-one correspondence between the states in the 1-MC and the network state $N(t)$. In particular, if the molecule is in the $j$-th state of the 1-MC at time $t$, then $N(t) = \mathds{1}_j$ and $q(t) = \mathds{1}_j$. 

\subsubsection{Multinomial distributions} 
\label{sec:solution:multi}
We will make extensive use of multinomial distributions. Let $m = [m_1,m_2,...,m_N]$ be a vector of non-negative integers and $p = [p_1,p_2,...,p_N]$ be a real vector with $0 \leq p_j \leq 1$ and $\sum_{j = 1}^N p_j \leq 1$, the multinomial distribution ${\cal M}(m,M,p)$ is defined as:
\begin{align}
{\cal M}(m,M,p) = \left\{ 
\begin{array}{cl}
M! \frac{ (1-|p|)^{(M-|m|}) }{ (M-|m|)! } \Pi_{j = 1}^N \frac{p_i^{m_j}}{m_j !} & \mbox{if } |m| \leq M \\
0 & \mbox{otherwise} 
\end{array}
\right.
\end{align} 
where $|p|$ and $|m|$ are respectively the 1-norm of $p$ and $x$. Note that in the above expression, we assume that $0^0 = 1$. 

A standard result that we will make use of is that the marginalization of a multinomial distribution is again a multinomial distribution. Let $\tilde{m} = [m_{k+1},...,m_N]$ and $\tilde{p} = [p_{k+1},...,p_N]$, then the marginalization of ${\cal M}(m,M,p)$ over the first $k$ random variables is:
\begin{align}
\sum_{m_1,m_2,\ldots,m_k} {\cal M}(m,M,p) = {\cal M}(\tilde{m},M,\tilde{p})
\end{align}
Finally, note that if $\tilde{m}$ is a scalar, the distribution is in fact binomial. 

\subsection{State probability distribution}  
We have the following theorem on the probability distribution of the state $N(t)$.  

\begin{thm}
Consider a network with a transmitter and a RC receiver. The initial network state $N(t)$ is zero. The transmitter, located at voxel $T$, emits $k_i$ molecules at time $t_i$ where $t_i > 0$ and $i = 1,2,3...$ . The probability that the state vector $N(t) = \nu$ is: 
\begin{align}
{\cal M}(\nu,k_{1},q(t-t_{1})) \Asterisk {\cal M}(\nu,k_{2},q(t-t_{2})) \Asterisk {\cal M}(\nu,k_{3},q(t-t_{3})) \Asterisk \ldots 
\label{eqn:conv} 
\end{align} 
where $\Asterisk$ denotes convolution. For $\tau \geq 0$, $q(\tau)$ is the state probability of the associated 1-MC and is the solution to: 
\begin{align}
\dot{q}(\tau) = H q(\tau)   \mbox{ with } q(0) = \mathds{1}_{T},  
\label{eq:mc1t1m} 
\end{align}
where $H$ is the generator of the 1-MC; if  $\tau < 0$, then $q(\tau) = 0$. 
$\Box$ 
\label{thm:rc} 
\end{thm} 

\noindent{\bf Proof:} 
We will establish the result step by step. Let us assume that the transmitter emits only once, and this takes place at time $t_1$ and the number of molecules emitted is $k_{1}$. Since the network state is zero initially, it means the network state remains zero until time $t_1$. At this time, $N(t_1) = k_{1} \mathds{1}_{T}$ which means $k_{1}$ molecules in the transmitter voxel. Since the network can be viewed as a collection of monomolecular reactions, we can now apply \cite[Proposition 1]{Jahnke:2007th} to show that the network state $N(t) = \nu$ is ${\cal M}(\nu,k_{1},q(t-t_{1}))$ where $q(\cdot)$ is the solution to \eqref{eq:mc1t1m}. Note that the time shift is needed in the multinomial distribution because the signalling molecules are emitted at time $t_1$. 

Let us now assume that the transmitter emits only twice: $k_1$ and $k_2$ molecules at, respectively, time $t_1$ and $t_2$. Since the network consists of a collection of monomolecular reactions, the molecules in the network evolves independently. We can consider the $k_{1}$ molecules emitted at time $t_{1}$ separately from those $k_{2}$ molecules emitted at time $t_{2}$. Let $N^{(1)}(t)$ and $N^{(2)}(t)$ denote the network state of the molecules emitted at time $t_{1}$ and $t_{2}$. Following on from the result in the last paragraph, the probability distributions of these two state vectors are respectively ${\cal M}(\cdot,k_{1},q(t-t_{1}))$ and ${\cal M}(\cdot,k_{2},q(t-t_{2}))$. The network state $N(t)$ is the sum of two independent random variables $N^{(1)}(t) + N^{(2)}(t)$. Therefore, the probability distribution for $N(t)$ is the convolution of the distributions of $N^{(1)}(t)$ and $N^{(2)}(t)$. In other words, the probability that $N(t) = \nu$ is: 
\begin{align}
{\cal M}(\nu,k_{1},q(t-t_{1})) \Asterisk {\cal M}(\nu,k_{2},q(t-t_{2})) 
\end{align} 
Equation \eqref{eqn:conv} is obtained by repeating the above argument. $\Box$ 

We remark that: (1) Theorem \ref{thm:rc} can also be proved by substituting the result into the RDMEX \cite[Equation 4]{Chou:rdmex_nc} \cite[Equation 9]{Chou:rdmex_tnb}, but such proof would be tedious and un-intuitive; (2) The above results can be extended to non-zero initial condition, multiple transmitters, multiple receivers and non-deterministic emission patterns of transmitters.

\subsection{Results on the number of output molecules} 
\subsubsection{Probability distribution of the number of output molecules} 
\label{sec:rc:opd}
Let $n_{RC}(t)$ denote the number of output molecules in the RC receiver at time $t$. Note that $n_{RC}(t)$ is a random process. The probability distribution of $n_{RC}(t)$ can be obtained by marginalising over the first $N_v+1$ elements of $N(t)$ given in Theorem \ref{thm:rc}. Since marginalization of a multinomial distribution is again multinomial, and marginalization and convolution are interchangeable, we have the probability that $n_{RC}(t) = \nu$ is 
\begin{align}
{\cal M}(\nu,k_{1},q_R(t-t_{1})) \Asterisk {\cal M}(\nu,k_{2},q_R(t-t_{2})) \Asterisk {\cal M}(\nu,k_{3},q_R(t-t_{3})) \Asterisk \cdots
\label{eqn:conv:o} 
\end{align} 
where $q_R(t)$ is the element in $q(t)$ related to the output molecule, see \eqref{eqn:qvec}. From the discussion in Section \ref{sec:rc:bg:1mc}, we know that $q_R(t)$ is the probability that a molecule circulating in the network taking the form of an output molecule in the receiver. Note that the individual distributions in equation \eqref{eqn:conv:o} are in fact binomial. 

\subsubsection{Mean number of output molecules} 
\label{sec:rc_mean} 
By using probability generating functions, it can be shown that the mean number of output molecules $\bar{n}_{RC}(t)$ in the RC receiver is equal to $u(t) \Asterisk q(t)$ where $u(t) = \sum_{i = 1}^\infty k_i \delta(t - t_i)$ and $\delta(\cdot)$ is the Dirac delta function. Furthermore, we show in the Appendix \ref{app:rc} that the Laplace transform of $\bar{n}_{RC}(t)$ is:
\begin{align}
\bar{N}_{RC}(s) = \frac{\rho_{RC}(s) G_{RT}(s) U(s)}{1 + s G_{RR}(s) \rho_{RC}(s)}. 
\label{eqn:mean_n_cr} 
\end{align}
Here, $U(s)$ is the Laplace transform of the transmitter emission pattern $u(t)$ and $\rho_{RC}(s) = \frac{k_+}{s+k_-}$ is the transfer function of the RC reaction. 

The Laplace transforms $G_{RT}(s)$ and $G_{RR}(s)$ are related to a network with its receiver reaction mechanisms {\sl removed}. This is the same as assuming $k_+$ and $k_-$ are both zero. This means that a signalling molecule in the network can only diffuse from voxel to voxel, leave the medium or decay, but {\sl cannot} be converted to a complex. 

To explain what $G_{RT}(s)$ is, we consider a network with receiver reaction mechanism removed. We assume that there is only one signalling molecule in the network at time zero and it is located in the {\sl transmitter} voxel. Let $g_{RT}(t)$ denote the probability that this signalling molecule is in the receiver voxel $R$ at time $t$. The $G_{RT}(s)$ is the Laplace transform of $g_{RT}(t)$. In fact, we can view $G_{RT}(s)$ as the transfer function from the transmitter voxel to the receiver voxel, hence the subscript $RT$ in the notation. Furthermore, in the absence of receiver reaction mechanisms, the mean number of signalling molecules in the receiver voxel is $G_{RT}(s) U(s)$.

For $G_{RR}(s)$, we again consider a network with receiver reaction mechanism removed. We assume that there is only one signalling molecule in the network at time zero and it is located in the {\sl receiver} voxel. Let $g_{RR}(t)$ denote the probability that this signalling molecule is in the receiver voxel $R$ at time $t$. $G_{RR}(s)$ is the Laplace transform of $g_{RR}(t)$. 

\begin{figure}
\begin{center}
\includegraphics[page=6,trim=0cm 6cm 0cm 6cm ,clip=true, width=10cm]{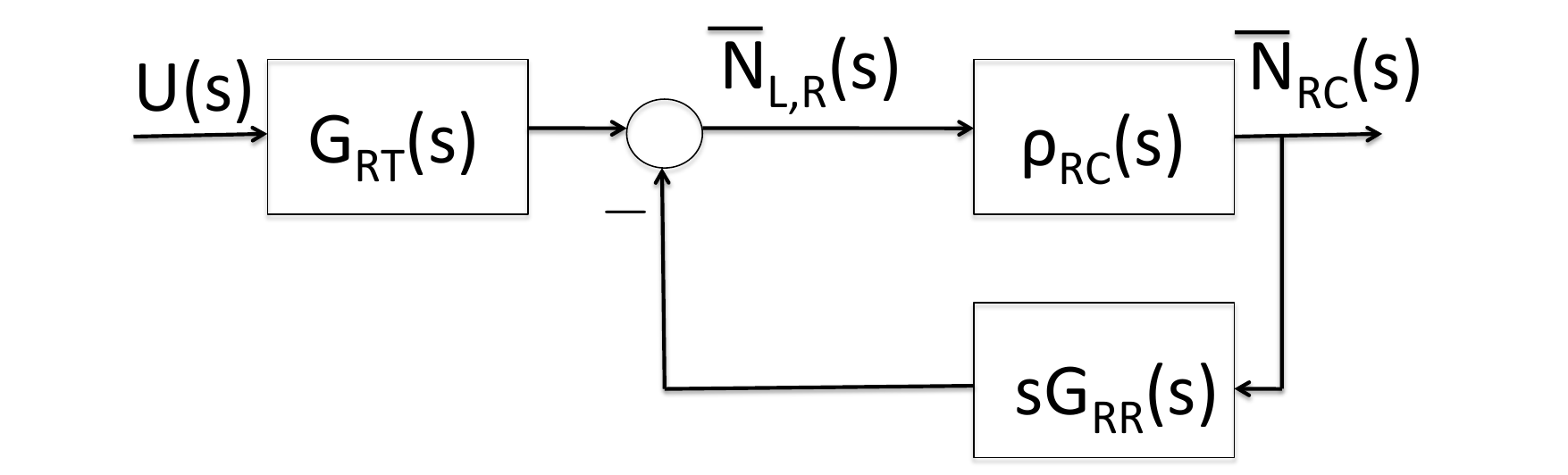}
\caption{Block diagram for the mean number of output molecules for RC receivers.}
\label{fig:block_rc}
\end{center}
\end{figure}

Equation \eqref{eqn:mean_n_cr} can be interpreted as a system with feedback, see Figure \ref{fig:block_rc}. Note that an expression similar to \eqref{eqn:mean_n_cr} has also been derived in our earlier work \cite{Chou:rdmex_tnb} assuming an homogeneous infinite transmission medium. 

We now present an example to demonstrate some of the properties of RC receivers. 

\begin{figure}
\begin{center}
\includegraphics[page=4,trim=0cm 11cm 10cm 2cm ,clip=true,width=10cm]{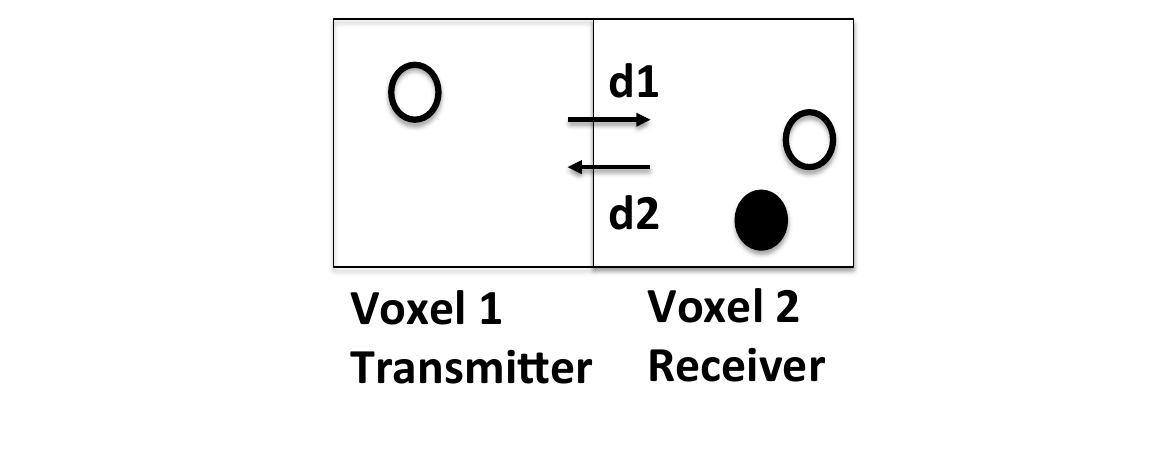}
\caption{Example network with 2 voxels.}
\label{fig:2voxel}
\end{center}
\end{figure}

\begin{figure}
\begin{center}
\includegraphics[page=5,trim=0cm 14cm 8cm 0cm ,clip=true, width=10cm]{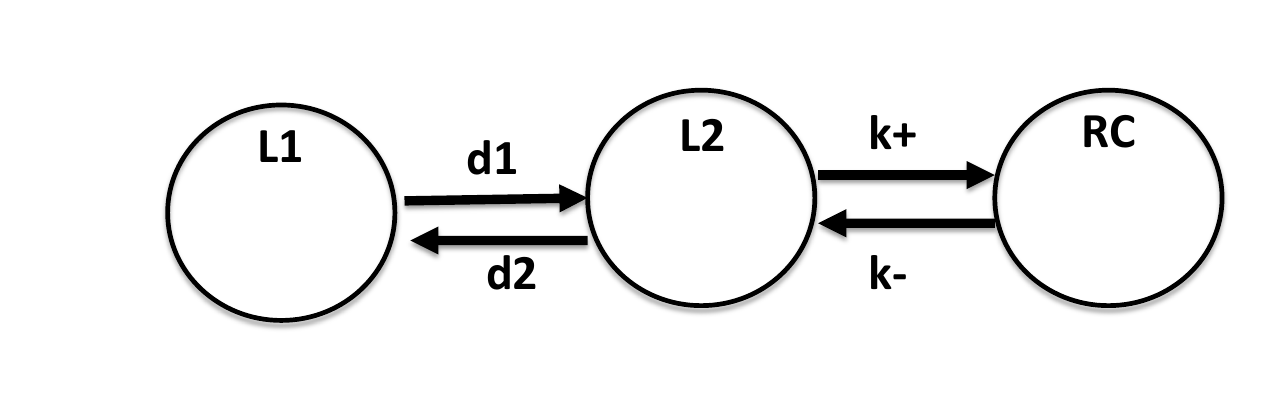}
\caption{The 1-MC for the model shown in Figure \ref{fig:2voxel}, assuming RC receivers.}
\label{fig:2voxel_1mc}
\end{center}
\end{figure}

\begin{example} 
Consider a molecular communication network with 2 voxels in Figure \ref{fig:2voxel}. Voxel 1 is the transmitter and voxel 2 is the receiver. The environment is heterogeneous where $d_1$ and $d_2$ are, respectively, the transition rates of signalling molecules leaving voxels 1 and 2. Here, we assume the receiver is of RC type. The corresponding 1-MC is shown in Figure \ref{fig:2voxel_1mc}. Let $q(t) = [q_{L,1}(t) ,  q_{L,2}(t) , q_{R}(t)]^T$ be the state probability vector. The generator $H$ of the 1-MC is:  
\begin{align}
H = 
\left[ \begin{array}{ccc}
-d_1 & d_2 & 0 \\
d_1 & -(d_2+k_+) & k_- \\
0 & k_+ & -k_-
\end{array}
\right]
\label{eqn:2voxel_RC_H}
\end{align} 
If the transmitter emits $N$ molecules intially and none afterwards. By solving $H q(\infty) = 0$,  the mean number of complexes $\bar{n}_{RC}$ (note: we drop $(\infty)$ for brevity) at steady state is: 
\begin{align}
\bar{n}_{RC} = \frac{N \frac{k_+}{k_-}}{1+\frac{k_+}{k_-}+\frac{d_2}{d_1}}
\label{eqn:2voxel_RC_nbar} 
\end{align}
The above result can also be derived by applying the Final Value Theorem to \eqref{eqn:mean_n_cr} and noting that as $\lim_{s \rightarrow 0}$, we have $s G_{RT}(s), s G_{RR}(s) \rightarrow \frac{d_1}{d_1 + d_2}$ and $\rho_{RC}(s) \rightarrow \frac{k_+}{k-}$. 

Equation \eqref{eqn:mean_n_cr} shows that one can increase the mean number of complexes by having a smaller $d_2$, i.e. making it harder for the signalling molecules to leave the receiver voxel. We will show in Section \ref{sec:num} that such heterogeneity in the permeability can be used to improve the performance of the receiver. 

The distribution of the number of output molecules is binomial. Therefore, the variance of the number of molecules $var(n_{RC})$ at steady state is $N \bar{n}_{RC} (1-\bar{n}_{RC})$. We have $var(n_{RC}) < \bar{n}_{RC}$ for RC receiver. $\Box$
\label{ex:rc}
\end{example}

\subsubsection{Variance and covariance} 
\label{sec:rc_var}
We can derive variance and covariance results from Theorem \ref{thm:rc}. For example, 
\begin{align}
var(n_{RC}(t)) & = u(t) \Asterisk (q_R(t) (1-q_R(t))) \label{eqn:rc_var_nrc} \\
cov(n_{L,i}(t),n_{L,j})(t) & = u(t) \Asterisk (-q_{L,i}(t) q_{L,j}(t)) \label{eqn:rc_covar_ex} 
\end{align} 
Equation \eqref{eqn:rc_var_nrc} gives the variance of the number of output molecules in the RC receiver. The covariance of the number of signalling molecules in voxels $i$ and $j$ is given in equation \eqref{eqn:rc_covar_ex}. These results can be derived by using the generating function of the probability distribution of $N(t)$. 

\section{CAT receivers} 
\label{sec:sol2}
This section presents results on the mean and variance of the number of output molecules in a CAT receiver. A special feature of the CAT reaction mechanism is that the signalling molecule $L$  appears on both sides of the chemical formula \eqref{cr:cat1}. This means that the dynamics of the signalling molecules is independent of that of the output molecules $E$. In fact, we can derive the probability distribution of the signalling molecule in the CAT network by using Theorem \ref{thm:rc} by removing the RC receiver. Let $\tilde{N}(t)$ and $\tilde{q}(t)$ be vectors containing the first $(N_v+1)$ elements of the state vector $N(t)$ (see \eqref{eqn:state}) and $q(t)$ (see \eqref{eqn:qvec}). In other words, $\tilde{N}(t)$ and $\tilde{q}(t)$ count only signalling molecules. 

\begin{figure}
\begin{center}
\includegraphics[width=10cm]{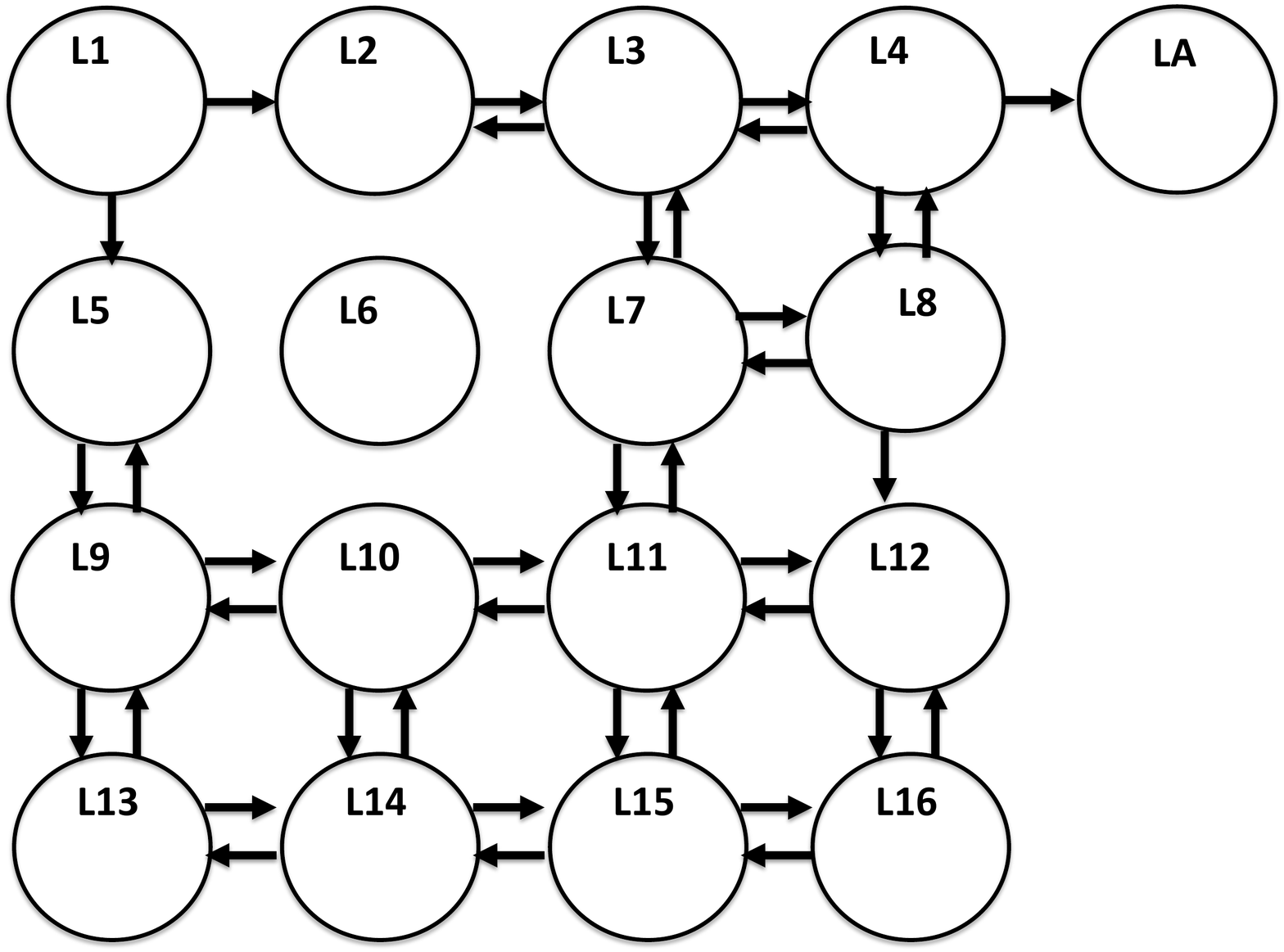}
\caption{The 1-MC for the model shown in Figure \ref{fig:model} with the receiver removed.}
\label{fig:mc_sm}
\end{center}
\end{figure}

Assuming that the transmitter emits $k_i$ molecules at time $t_i$ ($i = 1,2,...$), and the $\tilde{N}(0) = 0$. In a CAT network, we have the probability that  $\tilde{N}(t) = \nu$ is: 
\begin{align}
{\cal M}(\nu,k_{1},\tilde{q}(t-t_{1})) \Asterisk {\cal M}(\nu,k_{2},\tilde{q}(t-t_{2})) \Asterisk {\cal M}(\nu,k_{3},\tilde{q}(t-t_{3})) \Asterisk \ldots 
\label{eqn:cat_sm} 
\end{align} 
The vector $\tilde{q}(t)$ is the solution to 
\begin{align}
\dot{\tilde{q}}(\tau) = \tilde{H} \tilde{q}(\tau)   \mbox{ with } \tilde{q}(0) = \mathds{1}_{T} \mbox{ and } \tilde{q}(\tau) = 0 \mbox{ for } \tau < 0 
\label{eq:cat_sm_q} 
\end{align}
The matrix $\tilde{H}$ is the generator of the 1-MC with the receiver removed, i.e. the matrix $\tilde{H}$ accounts only for the diffusion of signalling molecules in the network. As an example, for the network in Figure \ref{fig:model}, we have drawn its 1-MC with its receiver removed in Figure \ref{fig:mc_sm}. The matrix $\tilde{H}$ is the generator for such types of 1-MC. 

Since we have the probability distribution of $\tilde{N}(t)$, we can show that the Laplace transform of the mean number of signalling molecules in the receiver voxel is $G_{RT}(s) U(s)$ where $G_{RT}(s)$ and $U(s)$ are as defined in Section \ref{sec:rc_mean}. In particular, $G_{RT}(s)$ is the Laplace transform of $g_{RT}(t)$, which is the probability that a signalling molecule emitted at the transmitter at time zero will be at the receiver voxel at time $t$. We can also obtain results on the variance and covariance of the number of signalling molecules in the voxels as in Section \ref{sec:rc_var}. 

\setcounter{subsubsection}{0}
\subsubsection{Mean number of output molecules} 
The following theorem is on the mean number of output molecules in the CAT receiver.  
\begin{thm}
Let $\bar{n}_{CAT}(t)$ denote the mean number of output molecules in the CAT receiver and $\bar{N}_{CAT}(s)$ be its Laplace transform. We have 
\begin{align}
\bar{N}_{CAT}(s) = \rho_{CAT}(s) G_{RT}(s) U(s)
\label{eqn:mean_n_cat} 
\end{align}
where $\rho_{CAT}(s) = \frac{f_+}{s + f_-}$ with $f_+$ and $f_-$ defined in Section \ref{sec:model}. 
$\Box$
\label{th:cat_mean} 
\end{thm} 

\begin{figure}
\begin{center}
\includegraphics[page=7,trim=0cm 9cm 0cm 6cm ,clip=true, width=10cm]{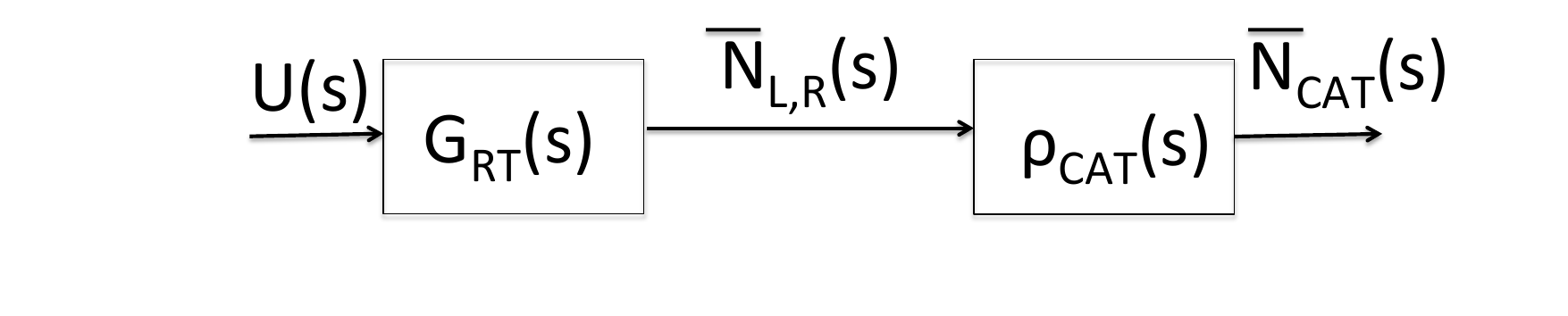}
\caption{Block diagram for the mean number of output molecules for CAT receivers.}
\label{fig:block_cat}
\end{center}
\end{figure}

The proof of this theorem can be found in Appendix \ref{app:cat}. Equation \eqref{eqn:mean_n_cat} can be depicted by the block diagram in Figure \ref{fig:block_cat}. The transmitter emission pattern $U(s)$ is first transformed by $G_{RT}(s)$ to obtain $\bar{N}_{L,R}(s)$ which is the Laplace transform of the mean number of signalling molecules in the receiver voxel. The latter is then subsequently transformed by $\rho_{CAT}(s)$ to obtain $\bar{N}_{CAT}(s)$. Although the transfer functions of RC and CAT receivers have the same form, their behaviours are different. In a RC receiver, signalling molecules are converted into complexes and this creates a feedback, see Figure \ref{fig:block_rc}. In a CAT receiver, the signalling molecule $L$  appears on both sides of the chemical formula \eqref{cr:cat1}, hence the absence of feedback. 


\subsubsection{Variance of the number of output molecules} The following theorem presents results on the variance of the number of output molecules. 

\begin{thm}
Let $var({n}_{CAT})(t)$ denote the variance of the number of output molecules in the CAT receiver and $var({N}_{CAT})(s)$ be its Laplace transform. We have 
\begin{align}
var({N}_{CAT})(s) = \frac{2}{s + 2 f_-} 
\left(
f_+^2 \mathds{1}_R^T \left(s I - \tilde{H} + k_- I \right)^{-1} \Psi(s) + f_+ \bar{N}_{L,R}(s) + f_- \bar{N}_{CAT}(s)
\right)
\label{eqn:var_n_cat} 
\end{align}
where $\Psi(s)$ is a $(N_v+1)$-dimensional vector and is the Laplace transform of the covariance vector $cov(\tilde{N}(t), n_{L,R}(t))$, i.e. the covariance between $\tilde{N}(t)$ and the number of signalling molecules $n_{L,R}(t)$ in the receiver voxel.
$\Box$
\label{th:cat_var} 
\end{thm} 

The proof of this theorem can be found in Appendix \ref{app:cat}. Note that $cov(\tilde{N}(t), n_{L,R}(t))$ can be computed because we know the distribution of $\tilde{N}(t)$. We will now use the 2-voxel network in Figure \ref{fig:2voxel} to illustrate some of the properties of the CAT receiver. 

\subsubsection{Two-voxel network example} 
\begin{example}
Consider the two voxel network in Figure \ref{fig:2voxel}. The receiver is now of the CAT type. The transmitter again emits $N$ molecules in the beginning. We will present expressions for steady state mean and variance of the number of output molecules. For convenience, we drop the $(\infty)$ for the time argument. The mean number of signalling molecules $\bar{n}_{L,R}$ in the receiver voxel is $\frac{N d_1}{d_1 + d_2}$. The mean number of output molecules $\bar{n}_{CAT}$ is 
\begin{align}
\bar{n}_{CAT} = \frac{f_+}{f_-} \bar{n}_{L,R} = \frac{f_+}{f_-} \frac{N d_1}{d_1 + d_2}. 
\label{eqn:2voxel_cat_ssm} 
\end{align}
By comparing \eqref{eqn:2voxel_RC_nbar}  and \eqref{eqn:2voxel_cat_ssm}, we can see that the CAT receiver has a higher mean output compared to the RC receiver if $f_+ = k_+$ and $f_- = k_-$. 

The steady state variance $\sigma_{CAT}^2$ of the number of output molecules is:
\begin{align}
\sigma_{CAT}^2 = \bar{n}_{CAT} + \frac{f_+^2}{f_-^2 + (d_1 + d_2) f_-} \sigma_{L,R}^2 
\label{eqn:ex:cat:var}
\end{align} 
where $\sigma_{L,R}^2$ is the variance of the number of signalling molecules in the receive voxel and is equal to $N \frac{d_1}{d_1 + d_2} \frac{d_2}{d_1 + d_2}$. The second term on the right-hand side of \eqref{eqn:ex:cat:var} deserves some explanation. The quantities $n_{L,R}(t)$ and $n_{CAT}(t)$ are related by a stochastic differential equation of the form:
\begin{align}
\dot{n}_{CAT}(t) = f_+ n_{L,R}(t) - f_- n_{CAT}(t) + \eta(t)
\end{align}
where $\eta(t)$ is a noise term. If $\eta(t)$ is independent of $n_{L,R}(t)$ and $n_{CAT}(t)$, then standard results on the response of linear time-invariant systems to noise \cite{Papoulis} says that the variance in $n_{CAT}(\infty)$ due to the variance of $n_{L,R}(\infty)$ is $(\frac{f_+}{f_-})^2 \sigma_{L,R}^2$. The noise amplification factor in \eqref{eqn:ex:cat:var} is in fact smaller. This is due to the fact that the noise $\eta(t)$ is correlated with both $n_{L,R}(t)$ and $n_{CAT}(t)$ in a chemical reaction, see \cite{Warren:2006ky}. 

Equation \eqref{eqn:2voxel_cat_ssm} shows that we can increase $\bar{n}_{CAT}$ by decreasing $d_2$, but this will also increase the variance $\sigma^2_{CAT}$. The mean output $\bar{n}_{CAT}$ can also be increased by adjusting the reaction constants $f_+$ and $f_-$ but a fundamental tradeoff is that an increase in mean also leads to an increase in variance, or noise. 

We will make a comparison between the RC and CAT receivers. Note that the same input has been used in Example \ref{ex:rc} and in this example. We further assume that $f_+ = k_+$ and $f_- = k_-$. Under these conditions, we have $\sigma_{CAT}^2 > \bar{n}_{CAT} > \bar{n}_{RC} > \sigma_{RC}^2$. We can view the mean and variance of the output as, respectively, the signal and noise strength. This means a RC receiver has a lower signal strength and a lower noise while a CAT receiver has a higher signal strength and a higher noise. We will further compare the performance of these two receivers in Section \ref{sec:num}. $\Box$ 
\label{ex:cat} 
\end{example}

\section{Numerical examples}
\label{sec:num} 
\begin{figure}
\begin{center}
\includegraphics[width=10cm]{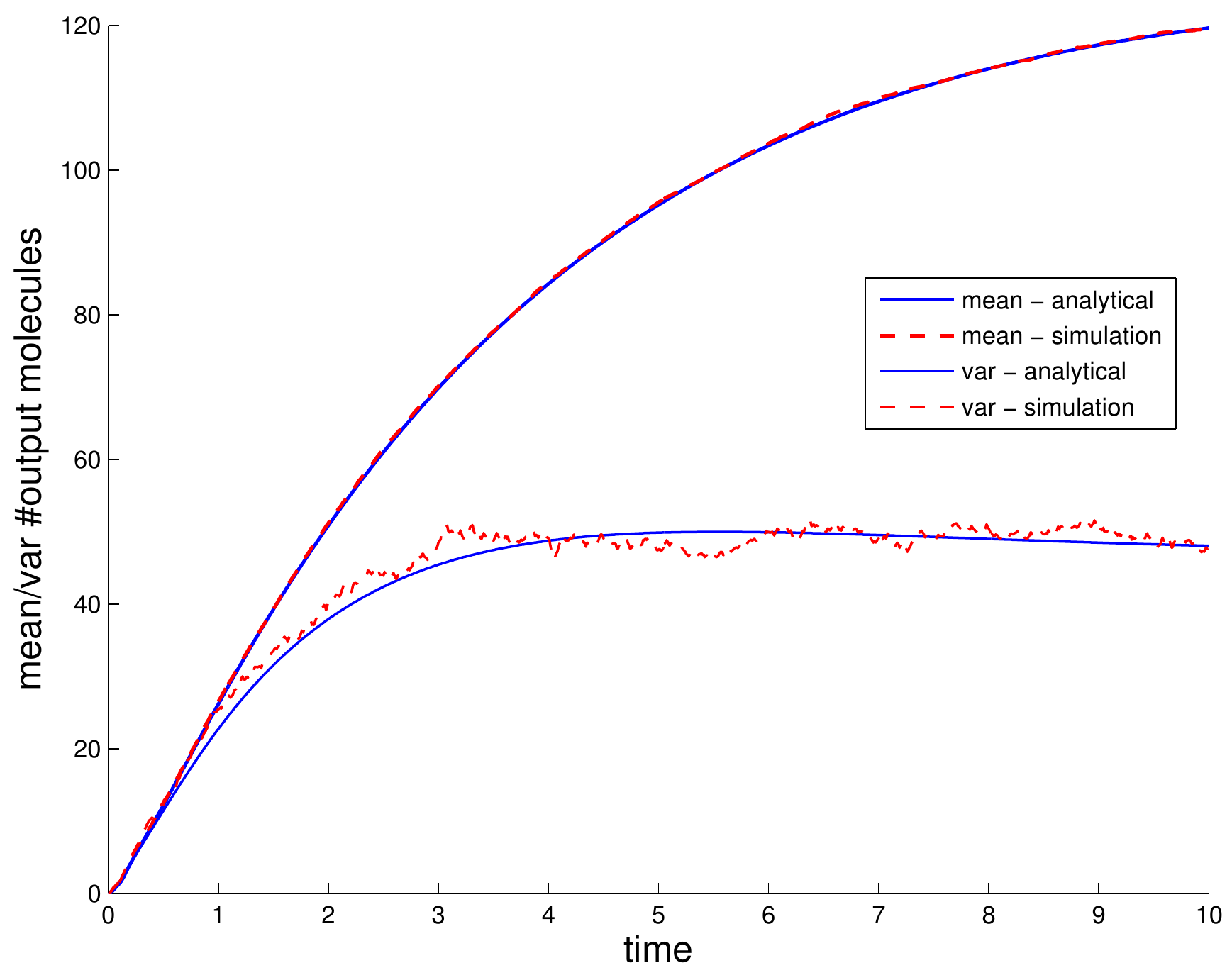}
\caption{Mean and variance of the number of output molecules in a RC receiver computed analytically and simulation.}
\label{fig:verify_rc}
\end{center}
\end{figure}

\begin{figure}
\begin{center}
\includegraphics[width=10cm]{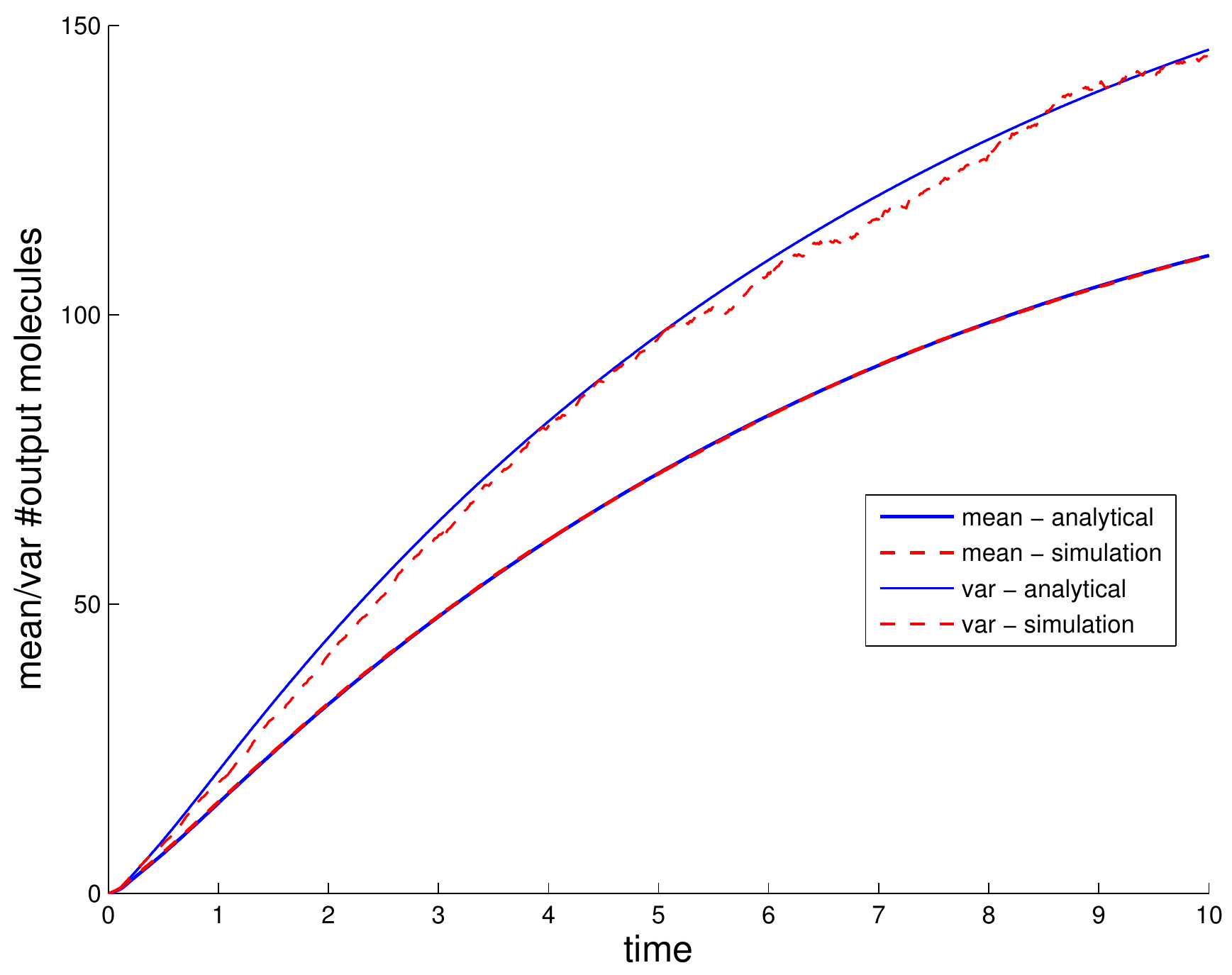}
\caption{Mean and variance of the number of output molecules in a CAT receiver computed analytically and simulation.}
\label{fig:verify_cat}
\end{center}
\end{figure}

In this section, we present numerical examples to illustrate the properties of the RC and CAT receivers. Before presenting the main findings, we first use a network of 75 voxels to verify the analytical solutions for computing the mean and variance of the number of output molecules. The results for RC and CAT receivers are presented in Figures \ref{fig:verify_rc} and \ref{fig:verify_cat} respectively. Each figure presents the mean and variance computed by analytical methods as well as by simulation. We used the $\tau$-leaping \cite{Gillespie:2001vc}, which is a commonly used method in chemistry to simulate diffusion and reactions of discrete molecules. The simulation results shown are the average of 1500 runs. The figures clearly show that the analytical expressions are accurate. 

\subsection{Mean and variance of the number of output molecules}

\subsubsection{Basic setup} 
We consider a medium of 10$\mu$m $\times$ 5 $\mu$m $\times$ 1 $\mu$m. We assume a voxel size of ($\frac{1}{3}$$\mu$m)$^{3}$ (i.e. $\Delta = \frac{1}{3}$ $\mu$m), creating an array of $30 \times 15 \times 3$ voxels. The transmitter and receiver are located at (2.5,2.5,0.5) and (7.5,2.5,0.5) (in $\mu$m) in the medium. The voxel co-ordinates are (7,7,3) and (22,7,3). We assume the diffusion coefficient $D$ of the medium is 1 $\mu$m$^2$s$^{-1}$. 

In all the comparisons below, we assume $K_+ = F_+$ and $k_- = f_-$, i.e. the RC and CAT receivers use the same parameters. We use two different values of $K_+$: 0.2 to 0.8 $\mu$m$^3$s$^{-1}$. The value of $k_-$ is 0.1 s$^{-1}$. These values are similar to those used in \cite{Erban:2009us} and are realistic for biological systems. We assume an absorbing boundary for the medium and the signalling molecules escape from the boundary voxel surface at a rate of $\frac{d}{500}$. The outflow permeability of the receiver voxel surface (i.e. the rate at which signalling molecules leave the receiver) is assumed to be $d$ unless otherwise stated. The transmitter emits at times 0, 1 and 2 seconds. The number of signalling molecules emitted at each time is 1000. 

\subsubsection{Mean of RC and CAT receivers} 
\begin{figure}
\begin{center}
\includegraphics[width=10cm]{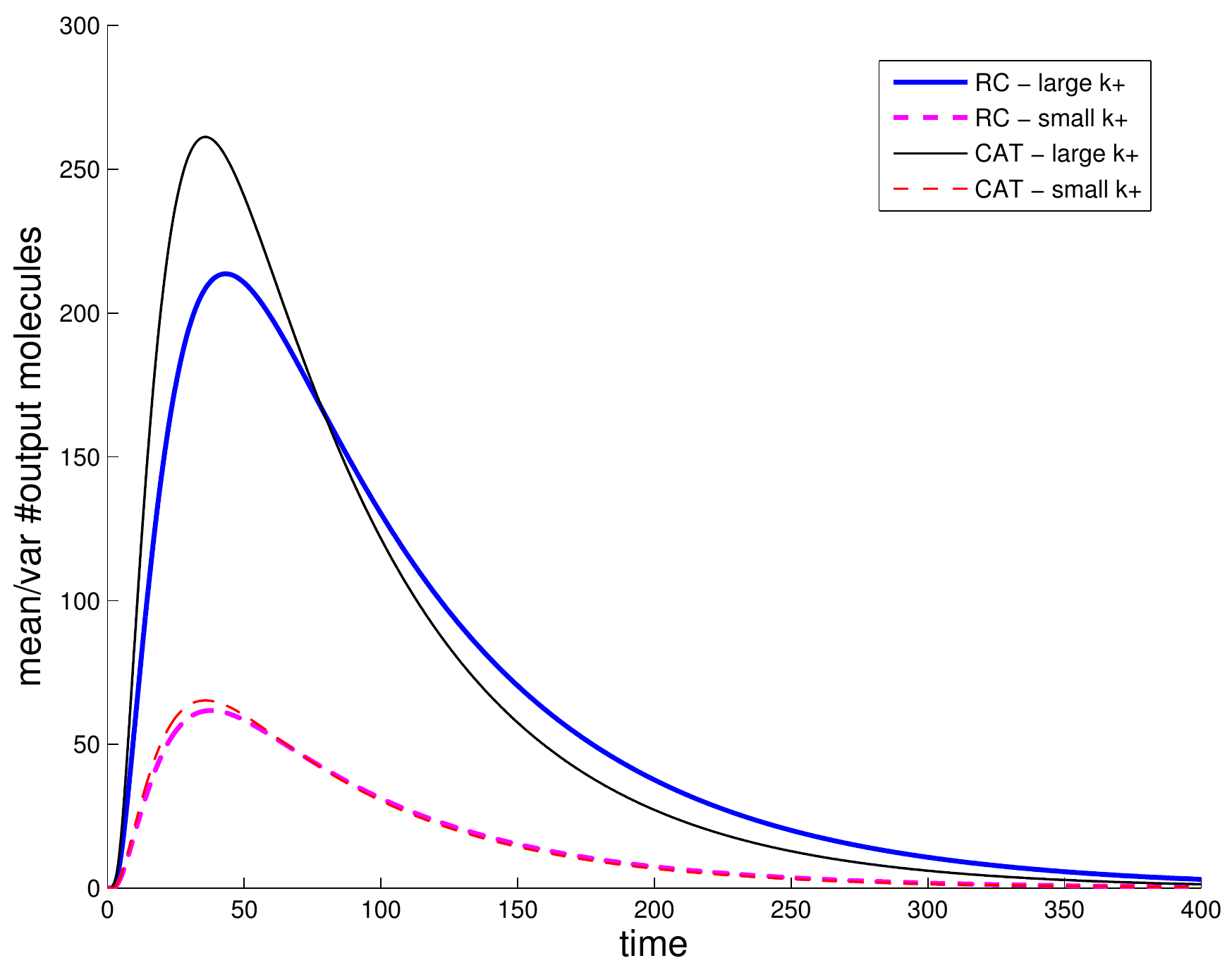}
\caption{Mean response of RC and CAT receivers for two different values of $k_+$.}
\label{fig:env2_mean_2k}
\end{center}
\end{figure}

Figure \ref{fig:env2_mean_2k} compares the mean output of RC and CAT receivers for two different values of $K_+ = $ 0.2 to 0.8. Two observations can be made. For smaller value of $K_+$, the mean output of both receiver types is almost the same. However, for larger value of $K_+$, the mean output of CAT is higher than that of RC. This can be explained by the form of transfer functions for RC and CAT receivers in equations \eqref{eqn:mean_n_cr} and \eqref{eqn:mean_n_cat} respectively. For small $K_+$, the denominator in \eqref{eqn:mean_n_cr} is almost equal to 1, so the two receivers have the same output. For large $K_+$, the denominator in \eqref{eqn:mean_n_cr} is large and results in RC having a lower mean output. 

\subsubsection{Mean and variance of RC and CAT receivers}
\begin{figure}
\begin{center}
\includegraphics[width=10cm]{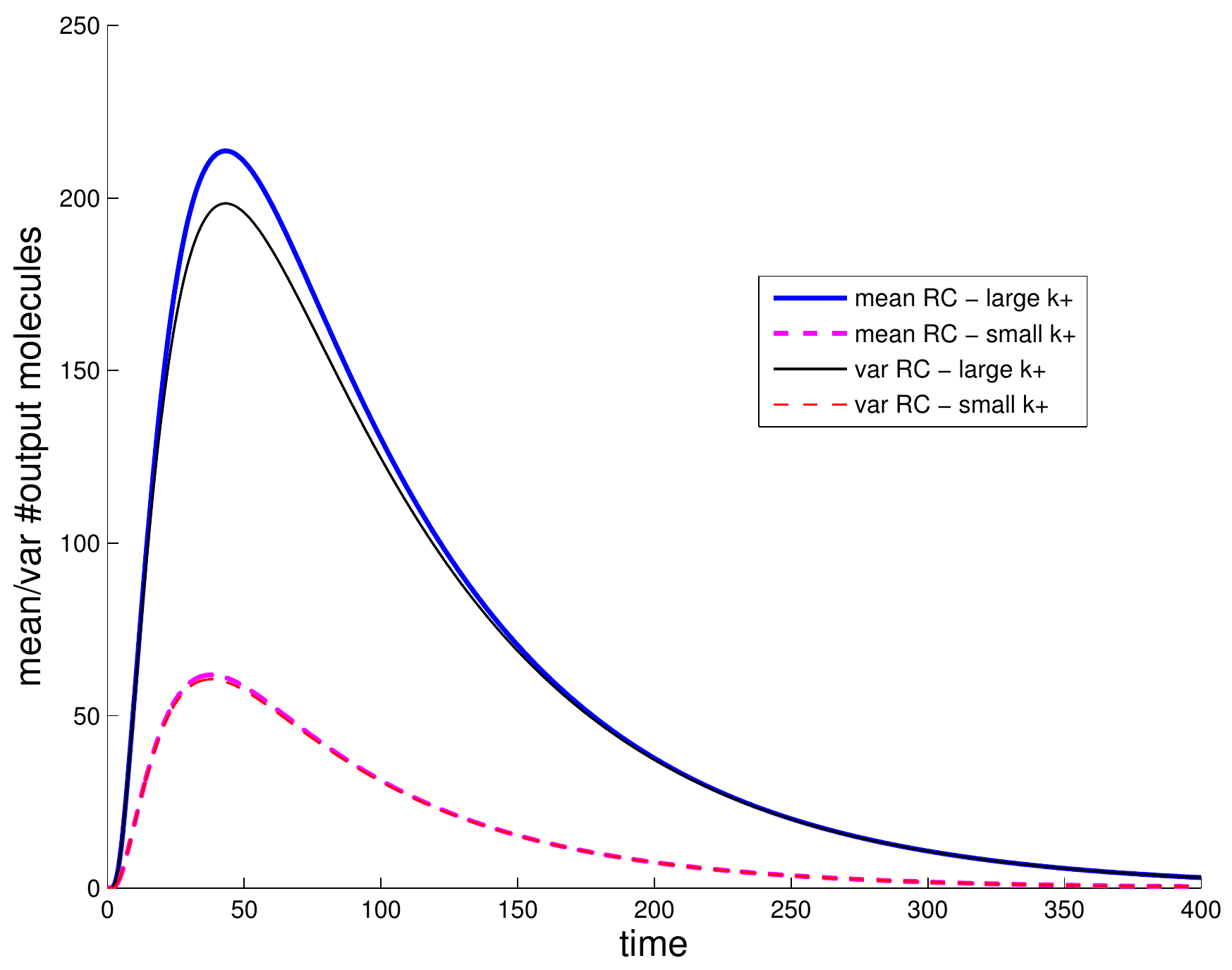}
\caption{Mean and variance of the RC receiver for two different values of $k_+$.}
\label{fig:rc_mv_2k}
\end{center}
\end{figure}

Figure \ref{fig:rc_mv_2k} shows for the mean and variance of the number of output molecules for the RC receiver for $K_+ = $ 0.2 to 0.8. It shows that for the RC receivers, the variance of the output signal is smaller than the mean of the signal. 

\begin{figure}
\begin{center}
\includegraphics[width=10cm]{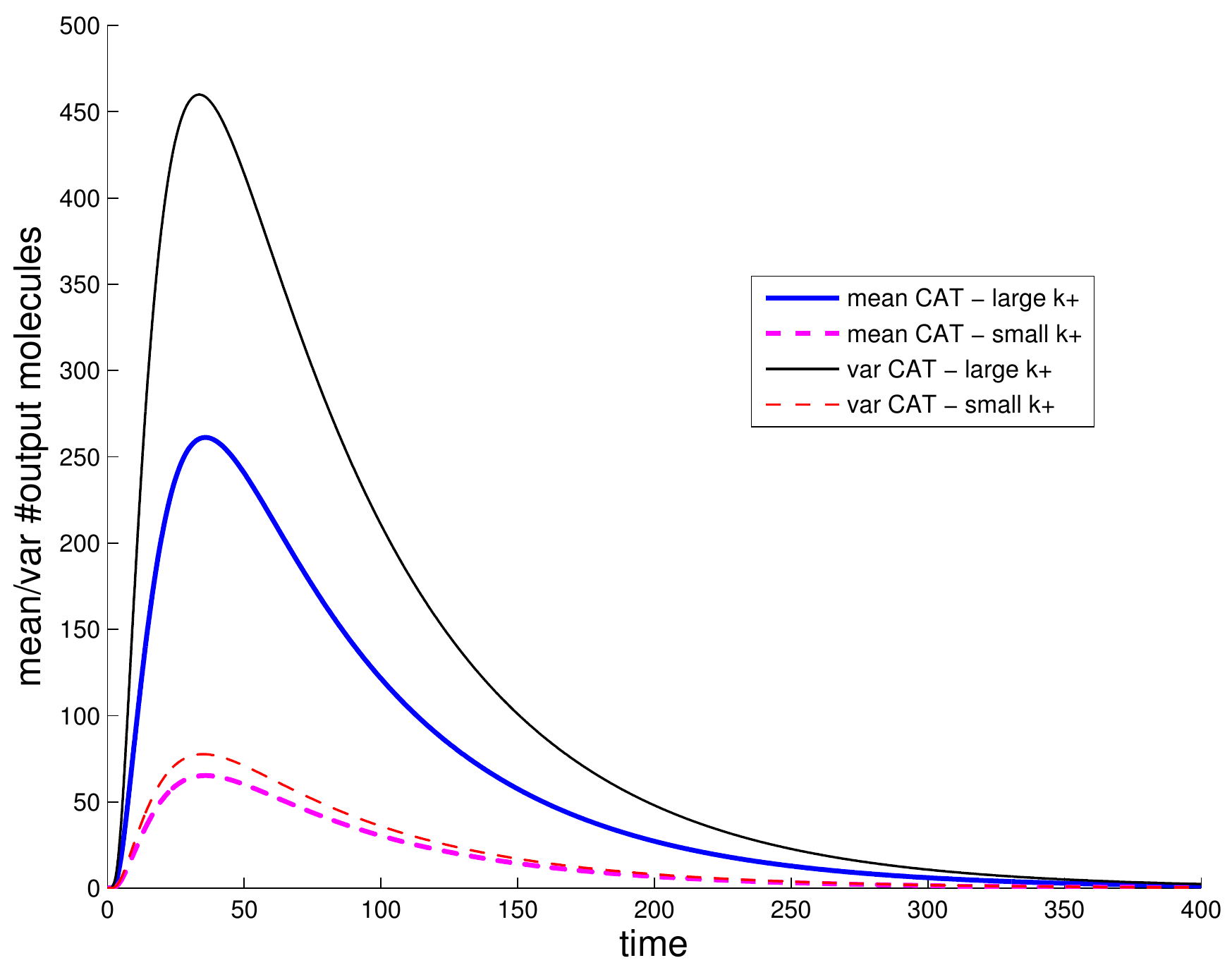}
\caption{Mean and variance of the CAT receiver for two different values of $k_+$.}
\label{fig:cat_mv_2k}
\end{center}
\end{figure}

Figure \ref{fig:cat_mv_2k} shows for mean and variance of the number of output molecules for the CAT receiver for $K_+ = $ 0.2 to 0.8. For CAT receivers, the variance of the receiver output is higher than the mean because CAT magnifies the fluctuations in the number of signalling molecules as discussed in Example \ref{ex:cat}. 

\subsubsection{Effect of receiver outflow permeability}
\begin{figure}
\begin{center}
\includegraphics[width=10cm]{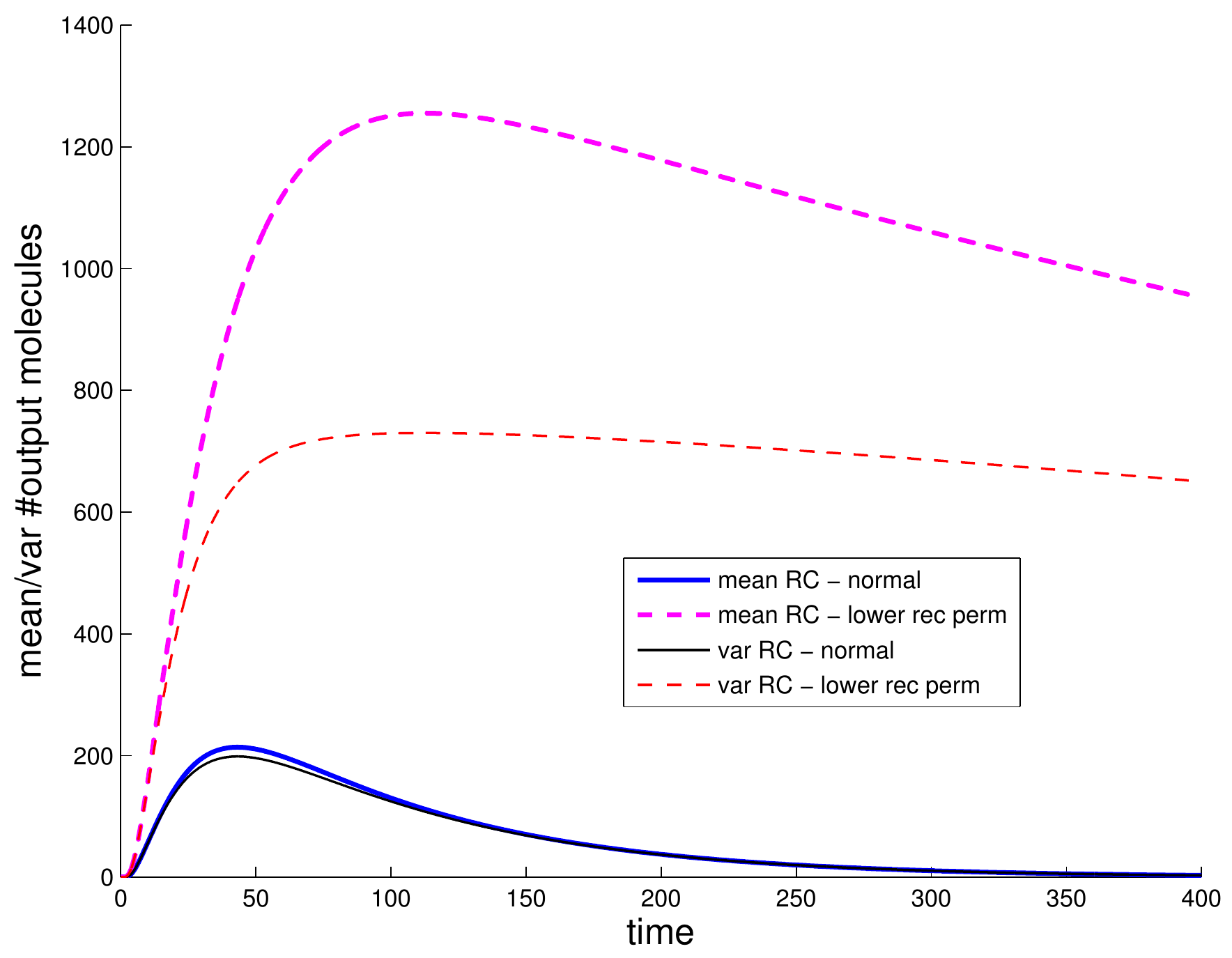}
\caption{Mean and variance of the RC receiver for two different outflow permeabilities at the receiver surfaces.}
\label{fig:rc_mv_2p}
\end{center}
\end{figure}

We investigate the impact of receiver outflow permeability on the mean and variance of the output signal. For the RC receivers, we compare the effect of a lower outflow permeability of  $\frac{d}{50}$ against the normal value of $d$. The lower value means signalling molecules leave the receiver at a lower rate of $\frac{d}{50}$. We keep $K_+ = 0.2$ in this study. Figure \ref{fig:rc_mv_2p} shows the mean and variance of the number of output molecules for the normal and lower permeabilities. It can be seen that a lower permeability magnifies the mean, and consequently, also the variance of the output signal. 

\begin{figure}
\begin{center}
\includegraphics[width=10cm]{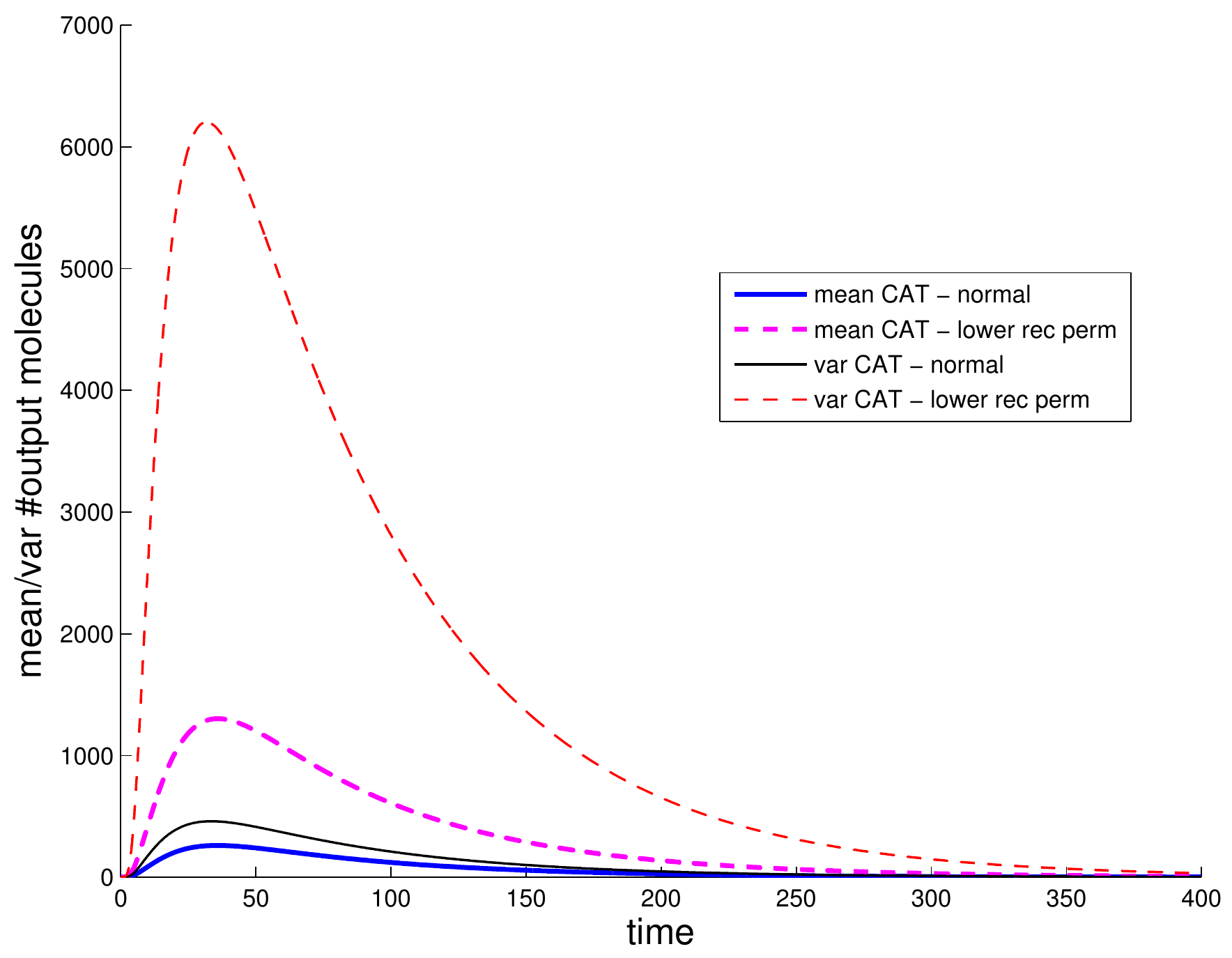}
\caption{Mean and variance of the CAT receiver for two different receiver outflow permeabilities at the receiver surfaces.}
\label{fig:cat_mv_2p}
\end{center}
\end{figure}

For CAT receivers, we compare the effect of a lower outflow permeability of $\frac{d}{5}$ against the normal value of $d$. Figure \ref{fig:cat_mv_2p} shows the mean and variance of the output signals of the CAT receiver. The same observations can be made: A lower receiver outflow permeability increases both the mean and variance of the output. If we use the same lower value of receiver outflow permeability, we find that the magnification in mean in variance for the CAT receivers is higher than that for the RC receivers. 

\subsubsection{Discussion}
We can draw a few observations from these studies. (1) Mean response for CAT receivers is higher than RC receivers if $K_+$ is large; otherwise they are almost the same. (2) The variance of RC receivers is always lower than the mean. (3) The variance of CAT receivers is higher than the mean. (4) The receiver outflow permeability can be used to increase the mean response of the receivers. 

\subsection{Capacity of memoryless channels} 
\label{sec:num:cap} 
We see in the above numerical results and in Example \ref{ex:cat} that for the same input signal and if $f_+ = g_+$ and $f_- = g_-$, then we have $\sigma_{CAT}^2(t) > \bar{n}_{CAT}(t) > \bar{n}_{RC}(t) > \sigma_{RC}^2(t)$. This means a RC receiver has a lower signal strength and a lower noise while a CAT receiver has a higher signal strength and a higher noise. It is not obvious which receiver has a better communication performance. In this section, we study the communication capacity of the two receivers as discrete memoryless channels. 

We assume the transmitter encodes the different messages by emitting a different number of molecules. This is comparable to pulse amplitude modulation. We assume that the transmitter can emit 10, 20, 30, ..., 1000 molecules at a time. This gives 100 input symbols. We use the steady state number of output molecules as the receiver output and we want to see how many different input levels the receiver can distinguish. We do that by using the Blahut algorithm \cite{Blahut:1972fw} to calculate the capacity of the discrete memoryless channel. We assume that consecutive symbols are well separated in time so that inter-symbol interference can be neglected. In order to use the Blahut algorithm, we need the probability distribution of the number of output molecules for each input symbol. This is available for the RC receiver but not for the CAT receiver. For the CAT receiver, we use the mean and variance of the output and assume that the probability is Gaussian distributed. Since Gaussian distribution has the maximum entropy, the estimated capacity is a lower bound of the true capacity \cite{Mitra:2001ib}. 

\begin{figure}[h] 
\begin{center}
\includegraphics[width=10cm]{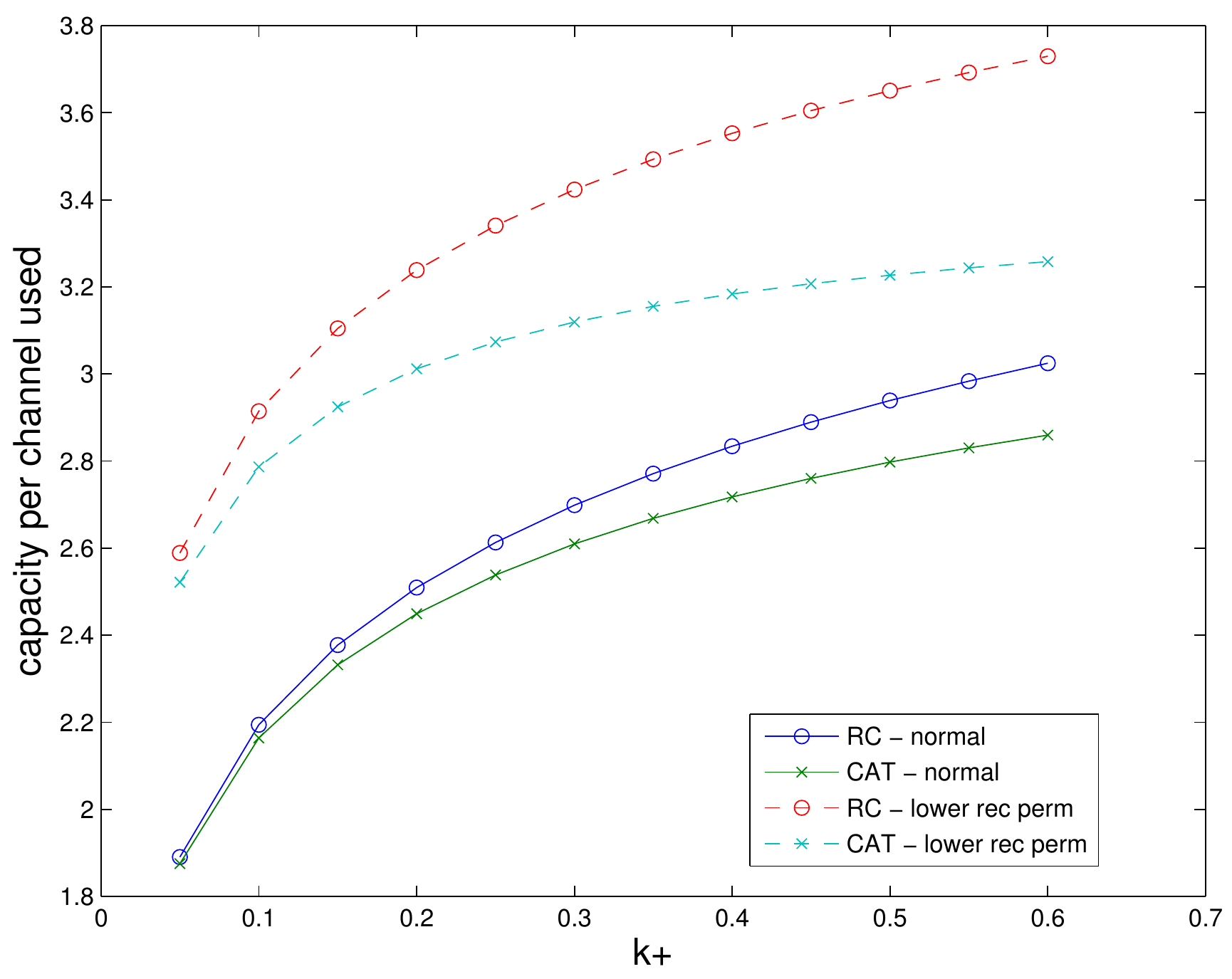}
\caption{Capacity per channel use versus $K_+$ for RC and CAT receivers under two different receiver outflow permeabilities.}
\label{fig:cap_kp}
\end{center}
\end{figure}

We assume a reflecting boundary in this study. This means the molecules cannot leave the medium. We use two different values of receiver outflow permeability: normal value of $d$ and a lower value of $\frac{d}{5}$. We vary the $K_+$ value from 0.05 to 0.6. Figure \ref{fig:cap_kp} shows the capacity for both receivers under two different receiver outflow permeabilities. For small $K_+$, RC and CAT receivers have almost the same capacity. However, for larger $K_+$, RC has a higher capacity compared to CAT for both receiver outflow permeabilities. This is due to CAT having a higher level of noise. 

We have observed earlier that a lower receiver outflow permeability can lead to a higher mean and variance of the output signals for both the RC and CAT receivers. Figure \ref{fig:cap_kp} shows that a lower receiver outflow permeability can lead to a higher channel capacity. 

\section{Related work} 
\label{sec:related}
Molecular communication plays a fundamental role in living organisms and have been widely studied in biology \cite{Alberts}. The study of molecular communication in the communication theory literature has been growing in the past decade. For recent review of this area, see \cite{Akyildiz:2008vt,Hiyama:2010jf,Nakano:2012dv}. Molecules in a molecular communication network can be propagated by active transport or diffusion. The former class of networks has been studied in \cite{Eckford:eq,Moore:2009eu} while the majority of the work assumes that molecules diffuse freely in the medium. This paper also assumes diffusion. 

Diffusion is a major source of noise in molecular communication networks. The work in \cite{Srinivas:2012et} considers using the emission time of a molecule for encoding. It shows that diffusion causes the deviation from expected arrival time to have an inverse Gaussian distribution. The authors of \cite{ShahMohammadian:2012iu} derive the probability distribution of the number of molecules arriving at a time at the molecular receiver and use the distribution to design a decoder. Both pieces of work consider only diffusion and do not consider the reaction mechanisms at the receiver. 

The authors in \cite{Pierobon:2011vr,Pierobon:2011ve} model the noise at the molecular receivers by tracking the particle dynamics of the molecules. The authors in \cite{Noel:2013ua, Noel:2013tr} assumes the receiver consists of a Michaelis-Menten catalytic process and shows that the receiver noise has a Poisson distribution. In our earlier work in \cite{Chou:rdmex_tnb,Chou:rdmex_nc}, we propose an alternative approach of using Master equations to study the noise in molecular communication network in a homogeneous infinite medium. In this work, we consider a finite medium and study the impact of heterogeneity on the performance of molecular receivers. The work in \cite{Chou:rdmex_nc} presents results on the mean and variance of the number of output molecules in a RC receiver. In this work, we derive its probability distribution. 

Receiver design is an important topic in communication theory. There is much recent work on decoder design for molecular communication, see \cite{Noel:2013tr,Chou:2012ug,ShahMohammadian:2013jm}. These pieces of work present decoders that can be used to decode a received signal. The receiver reaction mechanisms in these papers have been chosen beforehand. In this work, we compare two receiver reaction mechanisms to characterise their performance in terms of mean and variance. 

The capacity of diffusion based molecular communication network has been studied in \cite{Atakan:2010bj,Pierobon:2013cl}. Both papers model diffusion by means of diffusion partial differential equation. This paper takes a different approach and models the network using voxels. This modelling approach allows us to vary the permeability of the membrane of the receiver and study its impact on the receiver performance. Such type of heterogeneity does not appear be have been considered in earlier work. 

Reaction mechanisms similar to RC and CAT have also been studied in biophysics literature \cite{Warren:2006ky}. However, the study considers only the reaction mechanism itself, and does not consider transmitters and diffusion.

\section{Conclusions and future work} 
\label{sec:con}
This paper investigates the impact of different receiver reaction mechanisms on the performance of molecular communication networks. Two receiver mechanisms, reversible conversion and linear catalytic, have been studied. We derive analytical expressions for the mean and variance of the output signals for these two reaction mechanisms. For reversible conversion, we have also derived the probability distribution of the output signal. We find that linear catalytic magnifies the noise and results in a noisier output signal. This ultimately leads to linear catalytic having a lower communication performance compared to reversible conversion. The analytical expressions that we have derived also allow us to study the impact of the diffusibility of the receiver membrane on the mean and variance of the receiver output signal. We find that a selective receiver membrane, which makes it harder for the signalling molecules to leave the receiver, can be used to improve the communication performance. The use of a selective membrane to improve communication performance does not appear to have studied before.

\appendix

\section{Proof of equation \eqref{eqn:mean_n_cr} } 
\label{app:rc} 
Given that $\bar{n}_{CR}(t) = u(t) \Asterisk q_R(t)$, we have $\bar{N}_{CR}(s) = U(s) Q_R(s)$ where $Q_R(s)$ is the Laplace transform of $q_R(t)$. We will now derive an expression for $Q_R(s)$. 

We first partition $q(t)$ in \eqref{eqn:qvec} as: 
\begin{align}
q(t) = 
\left[
\begin{array}{c|c}
\tilde{q}(t) & q_R(t)  
\end{array}
\right]^T 
\end{align}
where $\tilde{q}(t)$ contains all but the last element in $q(t)$. According to Theorem \ref{thm:rc}, $q(t)$ is the solution to 
\begin{align}
\dot{q}(t) = H q(t) \mbox{ with } q(0) = \mathds{1}_T
\label{eq:qdotH_app}
\end{align}
where $H$ is the generator of the corresponding 1-MC. The matrix $H$ has a specific block structure, which allows us to write \eqref{eq:qdotH_app} as:
\begin{align}
\left[
\begin{array}{c}
\dot{\tilde{q}}(t) \\ \dot{q}_R(t)  
\end{array}
\right] 
 = 
\left[
\begin{array}{c|c}
\tilde{H} - \mathds{1}_R \mathds{1}_R^{T} k_+   &  \mathds{1}_R k_- \\ \hline 
\mathds{1}_R^T k_+ & - k_- 
\end{array}
\right] 
\left[
\begin{array}{c}
\tilde{q}(t) \\ q_R(t)  
\end{array}
\right] \mbox{ with } \tilde{q}(0) = \mathds{1}_T \mbox{ and } q_R(0) = 0 
\label{eqn:Hblock2} 
\end{align}
where $R$ is the index for the receiver voxel and $\mathds{1}_R$ is the $R$-th standard basis vector. The matrix $\tilde{H}$ contains only diffusion parameters and does not contain any parameters from the receiver reaction mechanism. In fact, $\tilde{H}$ is the generator of the 1-MC of a network with its receiver reaction mechanisms {\sl removed}. For example, for the network in Figure \ref{fig:model}, the 1-MC of the network with removed receiver is depicted in Figure \ref{fig:mc_sm}. 

By taking the Laplace transform of \eqref{eqn:Hblock2} and noting that $q_{L,R}(t) = \mathds{1}_R^T \tilde{q}(t)$ where $q_{L,R}(t)$ is the the probability that the signalling molecule is in the receiver voxel, we have, after some manipulations:
\begin{align}
Q_{L,R}(s) & =  
\underbrace{\mathds{1}_R^{T} (s I - \tilde{H})^{-1} \mathds{1}_T}_{= G_{RT}(s)} - 
\underbrace{\mathds{1}_R^{T} (s I - \tilde{H})^{-1} \mathds{1}_R}_{= G_{RR}(s)}
s Q_R(s) \\
s Q_R(s) & = k_+ Q_{L_R}(s) - k_- Q_R(s)
\end{align}
where $Q_{L,R}(s)$ and $I$ denote, respectively, the Laplace transform of $q_{L,R}(t)$ and the identity matrix. Equation \eqref{eqn:mean_n_cr} can now be obtained after eliminating $Q_{L,R}(s)$ from the above two equations. 

The transfer function $G_{RT}(s)$ is the Laplace transform of $g_{RT}(t)$ which is the solution of 
\begin{align}
\tilde{q}(t) = & \tilde{H} \tilde{q}(t) \mbox{ with } \tilde{q}(0) = \mathds{1}_T \nonumber \\ 
g_{RT}(t) = & \mathds{1}_R^T \tilde{q}(t) 
\end{align}
This means that $g_{RT}(t)$ is the probability that the signalling molecule is in the receiver voxel $R$ at time $t$ given that it is initially at the transmitter voxel. Since $\tilde{H}$ is used, this applies to the network with its receiver reaction mechanisms removed. The interpretation of $G_{RR}(s)$ is similar. 

\section{Proof of Theorems \ref{th:cat_mean} and \ref{th:cat_var}}
\label{app:cat} 
The state vector $N(t)$ of the CAT receiver network can be written as:
\begin{align}
N(t) = 
\left[ \begin{array}{c}
\tilde{N}(t) \\ \hline n_{CAT}(t)  
\end{array}
\right]
\end{align} 
where $\tilde{N}(t)$ is as defined in Section \ref{sec:sol2} and $n_{CAT}(t)$ is the number of output molecules in the CAT receiver at time $t$. We denote their means by $\langle \tilde{N}(t) \rangle$ and $\bar{n}_{CAT}(t)$. 

The CAT receiver network can be analysed by the RDMEX model in \cite{Chou:rdmex_tnb}, we have 
\begin{align}
\left[ \begin{array}{c}
\langle \dot{\tilde{N}} \rangle (t) \\ \hline  \dot{\bar{n}}_{CAT}(t)  
\end{array}
\right]
= 
\left[ \begin{array}{c|c}
\tilde{H} & 0 \\ \hline
\mathds{1}_R^T f_+ & f_- 
\end{array}
\right]
\left[ \begin{array}{c}
\tilde{n}(t) \\ \hline  \bar{n}_{CAT}(t)  
\end{array}
\right] + \mathds{1}_T \underbrace{\sum_{i=1}^\infty k_i \delta(t-t_i)}_{= u(t)} 
\label{eqn:rdemx_cat_mean} 
\end{align}
where $\tilde{H}$ is the generator matrix of the 1-MC with the receiver removed as discussed in Section \ref{sec:sol2}. 

By taking the Laplace transform of \eqref{eqn:rdemx_cat_mean} and noting that the Laplace transform of $u(t)$ is $U(s)$, $\rho_{CAT}(s) = \frac{f_+}{s + f_-}$ and $G_{RT}(s) = \mathds{1}_R^T (sI - \tilde{H})^{-1} \mathds{1}_T$, we can arrive at the expression of $\bar{N}_{CAT}(s)$ in the Theorem \ref{th:cat_mean}. 

Let $\Sigma$ be the covariance of the state vector $N(t)$. We partition $\Sigma$ into a 2$\times$2 block matrix conformal to the partitioning of $N(t)$ above. By applying the results in \cite{Chou:rdmex_tnb} on covariance of the RDMEX model to a CAT receiver network, the covariance matrix $\Sigma$ satisfies the following differential equation:  
\begin{align}
\left[ \begin{array}{c|c}
\dot{\Sigma}_{11} & \dot{\Sigma}_{12} \\ \hline
\dot{\Sigma}_{21} & \dot{\Sigma}_{22}
\end{array} \right]
 = & 
\left[ \begin{array}{c|c}
\tilde{H} & 0 \\ \hline
\mathds{1}_R^T f_+ & -f_- 
\end{array}
\right]
\left[ \begin{array}{c|c}
\Sigma_{11} & \Sigma_{12} \\ \hline
\Sigma_{21} & \Sigma_{22}
\end{array} \right]
+
\left[ \begin{array}{c|c}
\Sigma_{11} & \Sigma_{12} \\ \hline
\Sigma_{21} & \Sigma_{22}
\end{array} \right]
\left[ \begin{array}{c|c}
\tilde{H}^T & \mathds{1}_R f_+ \\ \hline
0 & -f_- 
\end{array} \right]
+ \nonumber \\
& 
\left[ \begin{array}{c|c}
* & 0 \\ \hline
0 & f_+ \bar{n}_{L,R}(t) + f_- \bar{n}_{CAT}(t) 
\end{array} \right]
\label{eqn:lyap}
\end{align}
where $*$ denotes a matrix whose details are not important for this derivation. 

Our goal is to obtain the variance $var(n_{CAT}(t))$ which is equal to $\Sigma_{22}$. Let $\Psi(t) = \Sigma_{11} \mathds{1}_R$ which is the $R$-th column of the covariance matrix $\Sigma_{11}$. Since $\Sigma$ is the covariance of the state vector $N(t)$, $\Sigma_{11}$ is the covariance of $\tilde{N}(t)$. Therefore the $R$-th column of $\Sigma_{11}$ is $cov(\tilde{N}(t) n_{L,R}(t))$. We can write the (1,2) and (2,2) blocks of \eqref{eqn:lyap} as
\begin{align}
\dot{\Sigma}_{12} & = (\tilde{H}-f_- I) \Sigma_{12} + f_+ \Psi(t) \\
\dot{\Sigma}_{22} & = -2 f_- \Sigma_{22} + 2 f_+ \mathds{1}_R^T \Sigma_{12} + 2 f_- \bar{n}_{CAT}(s) 
\end{align}
By taking the Laplace transform of the above two equations and after eliminating $\Sigma_{12}$, we arrive at the results of Theorem \ref{th:cat_var}.


\newpage
\listoffigures 
\newpage

%

%

\end{document}